\documentclass[11pt]{article}
\usepackage{amssymb}
\usepackage{amsmath}
\usepackage{natbib}
\usepackage{graphicx}
\usepackage{algorithm}
\usepackage{algpseudocode}
\usepackage{subfigure}
\usepackage{setspace}
\usepackage{amsthm}
\usepackage{bbm}
\usepackage{enumerate}
\usepackage{tabu}
\usepackage{palatino}
\usepackage{color}
\usepackage{array}
\usepackage{multirow}
\usepackage{verbatim}
\usepackage{float}

\usepackage[margin=2.8cm]{geometry}

\DeclareMathOperator{\sech}{sech}

\begin{document}

\setlength{\parindent}{0pc}
\setlength{\parskip}{1ex}

\date{\today}

\title{\textbf{Transformations in Semi-Parametric Bayesian Synthetic Likelihood}}

\author{Jacob W. Priddle$^{\dagger,\ddagger}$\footnote{Communicating Author: jacob.priddle@hdr.qut.edu.au} and Christopher Drovandi$^{\mathsection,\ddagger}$\\
\\
$^\ddagger$ School of Mathematical Sciences, Queensland University of Technology (QUT), Australia;\\
Australian Research Council Centre of Excellence for Mathematical and Statistical Frontiers;\\
QUT Centre for Data Science\\
$^\dagger$ ORCID ID: 0000-0003-1154-1139\\
$^\mathsection$ ORCID ID: 0000-0001-9222-8763
%$^\ddagger$ Australian Research Council Centre of Excellence for Mathematical and Statistical\\
% Frontiers (ACEMS)
}

\maketitle

\begin{center}\textbf{Abstract}\end{center}
Bayesian synthetic likelihood (BSL) is a popular method for performing approximate Bayesian inference when the likelihood function is intractable. In synthetic likelihood methods, the likelihood function is approximated parametrically via model simulations, and then standard likelihood-based techniques are used to perform inference. The Gaussian synthetic likelihood estimator has become ubiquitous in BSL literature, primarily for its simplicity and ease of implementation. However, it is often too restrictive and may lead to poor posterior approximations. Recently, a more flexible semi-parametric Bayesian synthetic likelihood (semiBSL) estimator has been introduced, which is significantly more robust to irregularly distributed summary statistics. In this work, we propose a number of extensions to semiBSL. First, we consider even more flexible estimators of the marginal distributions using transformation kernel density estimation. Second, we propose whitening semiBSL (wsemiBSL) -- a method to significantly improve the computational efficiency of semiBSL. wsemiBSL uses an approximate whitening transformation to decorrelate summary statistics at each algorithm iteration. The methods developed herein significantly improve the versatility and efficiency of BSL algorithms.\\

\textit{Keywords:} likelihood-free inference, approximate Bayesian computation (ABC), kernel density estimation, copula, covariance matrix estimation, Markov chain Monte Carlo. \\

\begin{comment}
Funding:\\
Conflicts of interest/Competing interests: none.\\
Availability of data and material: this article uses simulated data only.\\
Code availability: MATLAB code is provided for the methods introduced in this article.
\end{comment}

\newpage
%%%%%%%%%%%%%%%%%%%%%%%%%%%%%%%%%%%%%%%%%%%
%%%%%%%%%%%%%%%%%%%%%%%%%%%%%%%%%%%%%%%%%%%
\section{Introduction}
\label{sec:Introduction}

Simulator models are a type of stochastic model that is often used to approximate a real-life process. Unfortunately, the likelihood function for simulator models is generally computationally intractable, and so obtaining Bayesian inferences is challenging. Approximate Bayesian computation (ABC) \citep{sisson2018handbook} and Bayesian synthetic likelihood (BSL) \citep{price2018bayesian, wood2010statistical} are two methods for approximate Bayesian inference in this setting. Both methods eschew evaluation of the likelihood by repeatedly generating pseudo-observations from the simulator, given an input parameter value. ABC and BSL methods have been applied in many different fields; recently, in biology, to model the spread of the Banana Bunchy Top Virus \citep{varghese2020estimating}; in epidemiology, to model the transmission of HIV \citep{mckinley2018approximate} and tuberculosis \citep{lintusaari2019resolving}, and, in ecology, to model the dispersal of little owls \citep{hauenstein2019calibrating}. ABC is a more mature and established technique than BSL, and so it is more prevalent in applied fields. However, ABC can suffer from the curse of dimensionality with respect to the dimension of the summary statistic, requires a large number of model simulations, and the results can be highly dependent on a set of tuning parameters. BSL methods can be used to overcome many of these limitations. 
 
Synthetic likelihood methods approximate the likelihood function with a tractable distribution; in contrast, ABC methods are effectively non-parametric \citep{blum2010non}. The original synthetic likelihood method of \cite{wood2010statistical} approximates the summary statistic likelihood with a Gaussian distribution and then uses a Markov chain Monte Carlo (MCMC) sampler for maximum likelihood estimation. Later, \cite{price2018bayesian} consider the Gaussian synthetic likelihood in the Bayesian setting, and refer to their method as Bayesian synthetic likelihood. In practice, the Gaussian assumption of the summary statistic vector may be too restrictive, leading to a poor estimate of the likelihood, and then a poor estimate of the posterior. Herein, we refer to the Gaussian BSL method as standard BSL, denoted sBSL. 

A few authors have considered more flexible density estimators to improve the robustness of sBSL to irregular summary statistic distributions (e.g.\ \citealp{papamakarios2018sequential,an2020robust, fasiolo2018extended}). In particular, the semi-parametric Bayesian synthetic likelihood (semiBSL) method of \cite{an2020robust}, estimates the intractable summary statistic likelihood semi-parametrically -- non-parametrically estimating the marginal distributions using kernel density estimation (KDE), and parametrically estimating the dependence structure using a Gaussian copula. \cite{an2020robust} show empirically that semiBSL performs favourably to sBSL when summary statistics are distributed irregularly. semiBSL maintains many of the attractive properties of sBSL, including its scalability to a high dimensional summary statistic and ease of tuning.

Despite the appeal of semiBSL, the number of model simulations required to accurately estimate the correlation matrix scales poorly with the dimension of the summary statistic. The equivalent problem for sBSL, the scaling of the estimation of covariance matrix with the number of model simulations, has been explored by \cite{an2019accelerating}, \cite{ong2018likelihood}, \cite{ong2018variational}, \cite{everitt2017bootstrapped}, \cite{nott2019bayesian} and \cite{priddle2019efficient}. However, there are currently no methods designed specifically for the semi-parametric estimator, which, in practice, may preclude its application to problems where model simulation is computationally expensive. The first contribution of this article adapts and extends the methodology presented in \cite{priddle2019efficient}, which combines a whitening transformation with shrinkage covariance matrix estimation, to the semiBSL context.

SemiBSL provides additional robustness over sBSL when the summary statistic marginals deviate from normality. However, as we demonstrate in subsequent sections, for some distributions the KDE will fail. For instance, when a marginal summary statistic distribution has extremely heavy tails, the KDE will allocate essentially no density to the center of the distribution, and all weight to the tails (see Figure 2). In addition, it is well-known that the global bandwidth KDE rarely provides adequate smoothing over all features of the underlying distribution \citep{wand1991transformations, yang2003improved}. Our second contribution addresses this problem with a procedure that draws upon and extends the vast body of literature on density estimation. Specifically, we consider transformation kernel density estimation (TKDE, \citealp{wand1991transformations}) to estimate the marginal distributions of the summary statistic. The idea is to transform the distribution so that the standard global bandwidth KDE is accurate, and then transform back to the original domain to estimate the density. We adapt the hyperbolic power transformation of \cite{tsai2017hyperbolic}, and propose a procedure to effectively apply TKDEs in a semiBSL algorithm. 

The remainder of this article is structured as follows. In sections 2 and 3, we provide an overview of sBSL and semiBSL, respectively. In section 4, we present our method to significantly improve the computational efficiency of semiBSL. In section 5, we propose a new estimator of the marginal summary statistic distributions for semiBSL using TKDE. We assess the accuracy of the TKDEs on a number of test densities with known distribution. In section 6, we apply our new methods to four different examples. Last, we conclude in section 7.

\begin{comment}
applications. what types of applications is synthetic likelihood applied to. why is synthetic likelihood important

brief overview of how synthetic likelihood works

most research to date focuses on the Gaussian synthetic likelihood estimator
this is often too restric
\end{comment}

%%%%%%%%%%%%%%%%%%%%%%%%%%%%%%%%%%%%%%%%%%%
%%%%%%%%%%%%%%%%%%%%%%%%%%%%%%%%%%%%%%%%%%%

\section{BSL}
Synthetic likelihood algorithms are applicable in settings where the likelihood function $p(\boldsymbol{y}|\boldsymbol{\theta})$ is intractable but simulation from the model is straightforward, where $\boldsymbol{y} = (y_1,...,y_m)^\top$ (with $m\geq1$) is the set of observed data and $\boldsymbol{\theta}\in\boldsymbol{\Theta}\subset \mathbb{R}^p$ is an unknown parameter. Here, our target is the posterior distribution $p(\boldsymbol{\theta}|\boldsymbol{y})\propto p(\boldsymbol{y}|\boldsymbol{\theta})p(\boldsymbol{\theta})$, where $p(\boldsymbol{\theta})$ is the prior distribution on the parameter. 
In synthetic likelihood, among other likelihood-free algorithms, such as approximate Bayesian computation (ABC) (see \citealp{sisson+f18}), it is standard practice to degrade the data to a vector of informative summary statistics to help mitigate problems associated with dimensionality. Specifically, let $S(\cdot): \mathbb{R}^m\rightarrow\mathbb{R}^d$ be the summary statistic function that maps an $m$-dimensional dataset to a $d$-dimensional summary statistic. For $\boldsymbol{s}_{\boldsymbol{y}} = S(\boldsymbol{y})$, the implied target conditional on the summary statistic, often referred to as the partial posterior, is then $p(\boldsymbol{\theta}|\boldsymbol{s_y})\propto p(\boldsymbol{s_y}|\boldsymbol{\theta})p(\boldsymbol{\theta})$; depending (to a large extent) on the informativeness of the summary statistic, $p(\boldsymbol{\theta}|\boldsymbol{y})\approx p(\boldsymbol{\theta}|\boldsymbol{s}_y)$. However, since $p(\boldsymbol{y}|\boldsymbol{\theta})$ is intractable, it is generally the case that $p(\boldsymbol{s_y}|\boldsymbol{\theta})$ is also intractable, which leads us to consider sampling based methods that do not require evaluation of $p(\boldsymbol{s_y}|\boldsymbol{\theta})$ to obtain approximate inferences from the partial posterior.

In essence, synthetic likelihood methods assume a parametric form of the likelihood, which acts as a surrogate for the true likelihood and may be used directly in an MCMC (Markov chain Monte Carlo) sampler. In sBSL (see \citealp{price2018bayesian}), the summary statistic likelihood is approximated with a Gaussian distribution, $\mathcal{N}(\boldsymbol{s}_{\boldsymbol{y}}; \boldsymbol{\mu}(\boldsymbol{\theta}),\Sigma(\boldsymbol{\theta}))$. The synthetic likelihood parameters $\boldsymbol{\mu}(\boldsymbol{\theta})$ and $\boldsymbol{\Sigma}(\boldsymbol{\theta})$ are typically unknown, but a series of $n$ independent and identically distributed simulations from the model $\boldsymbol{x}_1,\dots,\boldsymbol{x}_n \sim p(\cdot|\boldsymbol{\theta})$ with corresponding summary statistics $S(\boldsymbol{x}_1),\dots,S(\boldsymbol{x}_n)$ can be used to construct the Monte Carlo estimates:
\begin{align}
    \boldsymbol{\mu}_n(\boldsymbol{\theta}) &= \frac{1}{n}\sum_{i=1}^nS(\boldsymbol{x}_i)\text{ and}\label{eq:samplemean}\\
    \boldsymbol{\Sigma}_n(\boldsymbol{\theta}) &= \frac{1}{n-1}\sum_{i=1}^{n}(S(\boldsymbol{x}_i)-\boldsymbol{\mu}_n(\boldsymbol{\theta}))(S(\boldsymbol{x}_i)-\boldsymbol{\mu}_n(\boldsymbol{\theta}))^\top\label{eq:samplecov}.
\end{align}
These may be used to yield the Gaussian synthetic likelihood estimator, $\mathcal{N}(\boldsymbol{s}_{\boldsymbol{y}}; \boldsymbol{\mu}_n(\boldsymbol{\theta}),\Sigma_n(\boldsymbol{\theta}))$, and the corresponding sBSL posterior approximation:
\begin{align*}
    p_{\text{sBSL}}(\boldsymbol{s}_y|\boldsymbol{\theta}) &= \int \mathcal{N}(\boldsymbol{s}_y|\boldsymbol{\mu}_n(\boldsymbol{\theta}),\boldsymbol{\Sigma}_n(\boldsymbol{\theta}))\prod_{i = 1}^n p(S(\boldsymbol{x}_i)|\boldsymbol{\theta})\ dS(\boldsymbol{x}_{1})\cdots S(\boldsymbol{x}_n)\\
    	p_{\text{sBSL}}(\boldsymbol{\theta}|\boldsymbol{s}_{\boldsymbol{y}}) &\propto p_{\text{BSL}}(\boldsymbol{s}_{\boldsymbol{y}}|\boldsymbol{\theta})p(\boldsymbol{\theta}).
\end{align*}

There are two main appeals of BSL: (1) that it can handle a relatively high dimensional summary statistic, and (2) that it can be more computationally efficient than competing likelihood-free Bayesian methods \citep{price2018bayesian, nott2019bayesian}. These are both direct benefits of specifying a parametric form of the summary statistic likelihood. However, as demonstrated by \cite{an2020robust}, in cases where the marginal summary statistic distributions deviate greatly from Gaussian, with, for example, heavy skewness, heavy tails or multiple modes, sBSL methods begin to break down. Often the posterior distribution will fail to adequately approximate the true partial posterior. In particularly challenging cases, the variance of the log synthetic likelihood estimator may be so large that the MCMC chain will become stuck within only a few iterations, and no discernible posterior distribution may be recovered (see Figure \ref{fig:stable_post}).

\section{semiBSL}
In this section, we provide an overview of the semiBSL method of \cite{an2020robust}. semiBSL provides additional robustness for a non-Gaussian distributed summary statistic. In semiBSL, the semi-parametric likelihood estimator is constructed as follows. Denote $\mathcal{S}^j$ the random variable corresponding to the $j^{\text{th}}$ summary statistic. Given the set of $n$ model simulations, the true PDF (probability density function) $g_{S^j}(s)$ is approximated using the kernel density estimate:
\begin{align}\label{eq:standard_kde}
\hat{g}_{S^j}(s) = \frac{1}{n}\sum_{i = 1}^n K_h(s-S(\boldsymbol{x}_i)^j),
\end{align}
where $K_h(u) = h^{-1}K(u/h)$ and $h$ is the bandwidth. The kernel function $K(\cdot)$ may be any symmetric PDF; in semiBSL, the Gaussian kernel $K(u) = 1/\sqrt{2\pi} \exp\left\{-u^2/2\right\}$ is used due to its simplicity and unbounded support. The above kernel density estimator uses a global (constant) bandwidth, selected according to the rule of \cite{silverman1986density}. It is straightforward to obtain the corresponding estimate of the CDF (cumulative density function) $\hat{G}_{S^j}(s)$ from the above equation.

Following estimation of the marginal summary statistic distributions, the dependence between the summaries is modelled via the Gaussian copula. Essentially, copula modelling allows the dependence structure and the marginal distributions to be estimated independently, allowing the user to consider alternative and more flexible marginal density estimators than the Gaussian distribution, as the case is in sBSL. For an introduction to copula models, we refer the reader to \cite{trivedi2007copula}. The Gaussian copula density, 
\begin{align*}
	c(\boldsymbol{u}) = \frac{1}{\sqrt{\text{det}(\boldsymbol{R})}}\exp\left\{-\frac{1}{2}\boldsymbol{\eta}^\top (\boldsymbol{R}^{-1}-\boldsymbol{I}_d)\boldsymbol{\eta}\right\}
\end{align*}
is parameterised by the correlation matrix $\boldsymbol{R}$ and the vector of standard Gaussian quantiles $\boldsymbol{\eta} = (\Phi^{-1}(u_1),\dots,\Phi^{-1}(u_d))^\top$, where $\Phi^{-1}(\cdot)$ is the inverse CDF of the standard normal distribution and $u_j = G_{S^j}(\boldsymbol{s}_{\boldsymbol{y}}^j)$ for $j = 1,\dots,d$. Replacing $G_{S^j}(s)$ with its kernel density estimate evaluated at the observed summary $\hat{G}_{S^j}(\boldsymbol{s}_{\boldsymbol{y}}^j)$, and $\boldsymbol{R}$ with the estimated correlation matrix $\boldsymbol{\hat{R}}$, we obtain the semiBSL posterior: 
\begin{align*}
	p_{\text{semiBSL}}(\boldsymbol{s_y}|\boldsymbol{\theta}) &= \int \frac{1}{\sqrt{\text{det}(\hat{\boldsymbol{R}})}} \exp\left\{-\frac{1}{2}\hat{\boldsymbol{\eta}}^\top_{\boldsymbol{s_y}}(\hat{\boldsymbol{R}}^{-1}-\boldsymbol{I}_d)\hat{\boldsymbol{\eta}}_{\boldsymbol{s_y}}\right\} \prod_{j=1}^d \hat{g}_j(s_{\boldsymbol{y}}^j) \prod_{i = 1}^n p(S(\boldsymbol{x}_i)|\boldsymbol{\theta})\ dS(\boldsymbol{x}_{1})\cdots S(\boldsymbol{x}_n)\\
	p_{\text{semiBSL}}(\boldsymbol{\theta}|\boldsymbol{s}_{\boldsymbol{y}}) &\propto p_{\text{semiBSL}}(\boldsymbol{s}_{\boldsymbol{y}}|\boldsymbol{\theta})p(\boldsymbol{\theta}).
\end{align*}
In the above equation, $\hat{\boldsymbol{\eta}}_{\boldsymbol{s_y}} = (\Phi^{-1}(\hat{u}_1),\dots,\Phi^{-1}(\hat{u}_d))^\top$ where $\hat{u}_j = \hat{G}_j(\boldsymbol{s}^j_{\boldsymbol{y}})$ for $j = 1,\dots,d$ and $\hat{\boldsymbol{R}}$ is estimated using a collection of $n$ simulated summary statistics $S(\boldsymbol{x}_1),\dots,S(\boldsymbol{x}_n)$. In practice, \cite{an2020robust} advocate to estimate $\boldsymbol{R}$ with the Gaussian rank correlation (GRC) (see \citealp{boudt2012gaussian}), which provides additional robustness to the potential lack of fit of the KDEs. 

We highlight two main limitations of semiBSL. First, the number of model simulations required to accurately estimate $\boldsymbol{R}$ scales poorly with $d$. This may be problematic for applications where model simulation is computationally expensive, especially if a relatively low dimensional and informative summary statistic is unavailable. Furthermore, the KDE is unreliable for distributions with extremely heavy tails, which may induce unduly high variance in the $p_{\text{semiBSL}}(\boldsymbol{s_y}|\boldsymbol{\theta})$ estimator and cause semiBSL to fail. In subsequent sections, we propose methods to overcome each of these limitations.

\section{Whitening semiBSL}\label{sec:wsemiBSL}
We now propose a method to improve the computational efficiency of semiBSL. Namely, we extend the whitening BSL (wBSL) methodology proposed by \cite{priddle2019efficient} to the semiBSL context. The motivation behind wBSL is articulated in Theorem 1 of \cite{priddle2019efficient}. The main consequence of the theorem is that for a Gaussian log synthetic likelihood estimator with diagonal covariance structure, $n$ must scale linearly with $d$ to control the variance of the estimator. On the other hand, to control the variance of the traditional Gaussian log synthetic likelihood estimator (that estimates the full covariance structure), $n$ must scale quadratically with $d$. This result suggests that there are significant computational benefits possible in BSL algorithms if the summary statistics are uncorrelated.   

Despite such a compelling result, it is a challenging problem to find a summary statistic vector that is both independent across its dimensions and retains a large proportion of the information content intrinsic to the observed data. The main idea of wBSL is that an approximate whitening or decorrelation transformation may be applied to the summary statistic at each algorithm iteration. In doing so, the covariance shrinkage estimator of \cite{warton2008penalized} may be applied with a high penalty, producing an accurate, low variance estimate of the likelihood function for a relatively small number of model simulations. If the full penalty is applied, this coincides with the Gaussian synthetic likelihood estimate with a diagonal covariance structure, and thus the desired computational gains may be achieved. In several empirical examples, \cite{priddle2019efficient} demonstrate that wBSL is able to produce an accurate partial posterior approximation, with an order of magnitude less model simulations than sBSL. Given the semi-parametric synthetic likelihood estimator uses the Gaussian copula, it is likely that it will inherit similar computational gains to the classical Gaussian estimator, particularly in cases where the marginal distributions are close to Gaussian. However, the extension of these concepts to semiBSL is not yet clear; here we provide an outline of our methodology, which we refer to as wsemiBSL. 

\begin{comment}
We propose a method to improve the computational efficiency of semiBSL. Namely, we extend the wBSL algorithm of \cite{priddle2019efficient} to the semi-parametric likelihood estimator. In wBSL, an approximate whitening transformation is applied to the summary statistics at each algorithm iteration, whereby the transformed summary statistics are approximately decorrelated with unit variance. In doing so, the covariance shrinkage estimator of \cite{warton2008penalized} may be applied with a high penalty, producing an accurate, low variance estimate of the likelihood function. In several empirical examples, \cite{priddle2019efficient} demonstrate that wBSL is able to produce accurate posterior approximations with an order of magnitude less model simulations than sBSL. Extending these concepts to the semiBSL context is challenging; here we provide an outline of our methodology.
\end{comment} 

Consider the Gaussian approximation of the summary statistic likelihood:
\begin{align*}
	\mathcal{N}(\boldsymbol{s_y};\boldsymbol{\mu},\boldsymbol{\Sigma}) = \frac{1}{\sqrt{(2\pi)^d\text{det}(\boldsymbol{\Sigma})}}\exp\left\{-\frac{1}{2}(\boldsymbol{s_y}-\boldsymbol{\mu})^\top\boldsymbol{\Sigma}^{-1}(\boldsymbol{s_y}-\boldsymbol{\mu})\right\},
\end{align*}
where the dependence of $\boldsymbol{\mu}$ and $\boldsymbol{\Sigma}$ on $\boldsymbol{\theta}$ has been suppressed for notational convenience. It is straightforward to show that:
\begin{align*}
	\mathcal{N}(\boldsymbol{s_y};\boldsymbol{\mu},\boldsymbol{\Sigma}) \propto \frac{1}{\sqrt{\text{det}(\boldsymbol{R})}}\exp\left\{-\frac{1}{2}\boldsymbol{\eta_{s_y}}^\top\boldsymbol{R}^{-1}\boldsymbol{\eta_{s_y}}\right\}\prod_{j=1}^d\frac{\mathcal{N}(\eta_y^j;0,\sigma_j^2)}{\phi(\hat{\eta}_y^j)},
\end{align*}
where $\boldsymbol{\Sigma} = \boldsymbol{\Sigma}_d^{1/2} \boldsymbol{R} \boldsymbol{\Sigma}_d^{1/2}$ and $\boldsymbol{\Sigma}_d = \text{diag}(\sigma_1^2,\dots,\sigma_d^2)$.

The main disparity between wBSL and wsemiBSL, is that in wsemiBSL the whitening transformation is applied to the standard Gaussian quantiles, and not directly to the summary statistics. We find that in the context of semiBSL, the latter approach does not produce as accurate posterior approximations (results not shown). Specifically, we propose to apply the whitening transformation to convert the random vector $\boldsymbol{\eta}$ of arbitrary distribution with covaraince matrix $\text{Var}(\boldsymbol{\eta}) = \boldsymbol{R}$ into the transformed vector
\begin{align*}
	\tilde{\boldsymbol{\eta}} = \boldsymbol{W}\boldsymbol{\eta}
\end{align*}
for some $d\times d$ whitening matrix $\boldsymbol{W}$, such that the covariance $\text{Var}(\tilde{\boldsymbol{\eta}}) = \boldsymbol{I}_d$ is the identity matrix. Like in wBSL, we estimate the whitening matrix off-line using $n_{\text{cov}}$ independent model simulations such that $\boldsymbol{x}_1,\dots,\boldsymbol{x}_{n_{\text{cov}}}\sim p(\cdot|\boldsymbol{\theta}^0)$ given some carefully chosen parameter value $\boldsymbol{\theta}^0$ with reasonable posterior support. \cite{picchini2020adaptive} detail how Bayesian optimization may be used to rapidly generate a $\boldsymbol{\theta}^0$ that has reasonable support under the posterior. This method, or the techniques described in \cite{priddle2019efficient}, may be employed to find a suitable $\boldsymbol{\theta}^0$ for our methods. $n_{\text{cov}}$ is set high (much higher than $n$) to ensure an accurate estimate of $\boldsymbol{W}$ is obtained. Of course, for the transformation to be exact, $\boldsymbol{W}$ must evolve as a function of $\boldsymbol{\theta}$; however, like in wBSL, we hold $\boldsymbol{W}$ constant to preserve the target partial posterior obtained using semiBSL (when no penalty is applied), and so generally $\text{Var}(\tilde{\boldsymbol{\eta}})\approx \boldsymbol{I}_d$. Given the inverse transformation $\boldsymbol{\eta} = \boldsymbol{W}^{-1}\tilde{\boldsymbol{\eta}}$ and Jacobian term $|d\boldsymbol{\eta}/d\tilde{\boldsymbol{\eta}}| = \text{det}(\boldsymbol{W}^{-1})$, the summary statistic likelihood under the transformed variable is
\begin{align*}
	\tilde{g}(\boldsymbol{s_y}|\boldsymbol{\theta})&\propto \frac{\text{det}(\boldsymbol{W}^{-1})}{\sqrt{\text{det}(\boldsymbol{R})}}\exp\left\{-\frac{1}{2}(\boldsymbol{W}^{-1}\tilde{\boldsymbol{\eta}}_{\boldsymbol{s_y}})^\top\boldsymbol{R}^{-1}\boldsymbol{W}^{-1}\tilde{\boldsymbol{\eta}}_{\boldsymbol{s_y}}\right\}
	\prod_{j=1}^d\frac{\mathcal{N}(\eta_{s_y}^j;0,\sigma_j^2)}{\phi(\eta_{s_y}^j)}\\
	&= \frac{1}{\sqrt{\text{det}(\tilde{\boldsymbol{\Sigma}}_{\boldsymbol{\eta}})}}\exp\left\{-\frac{1}{2}\tilde{\boldsymbol{\eta}}_{\boldsymbol{s_y}}^\top\tilde{\boldsymbol{\Sigma}}^{-1}_{\boldsymbol{\eta}}\tilde{\boldsymbol{\eta}}_{\boldsymbol{s_y}}\right\}
	\prod_{j=1}^d \frac{\mathcal{N}(\eta_{s_y}^j;0,\sigma_j^2)}{\phi(\eta_{s_y}^j)},
\end{align*}
where $\tilde{\boldsymbol{\Sigma}}_{\boldsymbol{\eta}} = \boldsymbol{WR}\boldsymbol{W}^\top = \text{Var}(\tilde{\boldsymbol{\eta}}_{\boldsymbol{s_y}})\approx \boldsymbol{I}_d$ is the covariance matrix of the transformed quantiles $\boldsymbol{\tilde{\eta}}_{\boldsymbol{s_y}}$. Of course, in semiBSL, we replace each marginal $\mathcal{N}(\eta_{s_y}^j;0,\sigma_j^2)$ with the kernel density estimate $\hat{g}_{S^j}(s)$ and $\tilde{\boldsymbol{\Sigma}}_{\boldsymbol{\eta}}$ with a sample estimate. That is, 
\begin{align*}
	\tilde{g}(\boldsymbol{s_y}|\boldsymbol{\theta}) \propto \frac{1}{\sqrt{\text{det}(\tilde{\boldsymbol{\Sigma}}_{\boldsymbol{\eta}})}} \exp\left\{ - \frac{1}{2}\hat{\tilde{\boldsymbol{\eta}}}_{\boldsymbol{s_y}}^\top \hat{\tilde{\boldsymbol{\Sigma}}}_{\boldsymbol{\eta}}^{-1}\hat{\tilde{\boldsymbol{\eta}}}_{\boldsymbol{s_y}} \right\}
	\prod_{j=1}^d \frac{\hat{g}_j(s_y^j)}{\phi(\hat{\eta}_{\boldsymbol{s_y}}^j)}.
\end{align*}
where $\hat{\tilde{\boldsymbol{\eta}}}_{\boldsymbol{s_y}} = \boldsymbol{W}(\Phi^{-1}(\hat{u}_1),\dots,\Phi^{-1}(\hat{u}_d))^\top$ and $\hat{u}_j = \hat{G}_j(\boldsymbol{s}^j_{\boldsymbol{y}})$ for $j = 1,\dots,d$. $\hat{\tilde{\boldsymbol{\Sigma}}}_{\boldsymbol{\eta}}$ is estimated using $n$ simulated quantiles $\hat{\tilde{\boldsymbol{\eta}}}_{S(\boldsymbol{x}_1)},\dots, \hat{\tilde{\boldsymbol{\eta}}}_{S(\boldsymbol{x}_n)}$ which constitute the rows of the $n\times d$ matrix $\boldsymbol{W}(\hat{\boldsymbol{\eta}}_{S(\boldsymbol{x}_1)},\dots,\hat{\boldsymbol{\eta}}_{S(\boldsymbol{x}_n)})^\top$ such that $\hat{\boldsymbol{\eta}}_{S(\boldsymbol{x}_i)} = (\Phi^{-1}(\hat{u}_i^1),\dots,\Phi^{-1}(\hat{u}_i^d))^\top$ and $\hat{u}_i^j = \hat{G}_j(S(\boldsymbol{x}_i)^j)$ for $j = 1,\dots,d$ and $i = 1,\dots,n$. Given the whitening transformation approximately decorrelates the summary statistic quantiles, the \cite{warton2008penalized} covariance shrinkage estimator
\begin{align*}
	\tilde{\boldsymbol{\Sigma}}_{\boldsymbol{\eta},\gamma} = \tilde{\boldsymbol{\Sigma}}^{1/2}_{\boldsymbol{\eta},d}(\gamma\tilde{\boldsymbol{R}}_{\boldsymbol{\eta}} + (1-\gamma)\boldsymbol{I}_d)\tilde{\boldsymbol{\Sigma}}^{1/2}_{\boldsymbol{\eta},d}
\end{align*}
may be applied accurately with a high degree of shrinkage, where $\tilde{\boldsymbol{\Sigma}}_{\boldsymbol{\eta},d} = \text{diag}(\tilde{\boldsymbol{\Sigma}}_{\boldsymbol{\eta}})$, $\tilde{\boldsymbol{R}}_{\boldsymbol{\eta}}$ is an estimate of the correlation matrix and $\gamma\in[0,1]$ is the shrinkage parameter. Effectively, $\gamma$ is a constant that is multiplied by the off-diagonal elements of the sample covariance. Thus, $\gamma = 0$ shrinks the pairwise covariance elements to $0$, assuming independent summary statistic quantiles. The heavier the shrinkage, the lower the value of $n$ required to precisely estimate the likelihood.

The choice of whitening matrix $\boldsymbol{W}$ was considered carefully in \cite{priddle2019efficient}. Any $\boldsymbol{W}$ that satisfies $\text{Var}(\tilde{\boldsymbol{\eta}}) = \text{Var}(\boldsymbol{W\eta}) = \boldsymbol{W\Sigma}\boldsymbol{W}^\top = \boldsymbol{I}_d$ will whiten the data at $\boldsymbol{\theta}^0$; however, as the current parameter value deviates further from $\boldsymbol{\theta}^0$, the transformation will become less accurate. The most suitable $\boldsymbol{W}$ for BSL is the one that most effectively decorrelates summary statistics generated by parameter values that reside in regions of the parameter space with non-negligible posterior density. \cite{priddle2019efficient} consider the five optimal whitening matrices of \cite{kessy2018optimal}. \cite{priddle2019efficient} find that in the context of BSL, principal components analysis (PCA) whitening produces the most accurate partial posterior approximations upon the application of heavy shrinkage. Thus, in wsemiBSL we also use the PCA whitening matrix,
\begin{align*}
	\boldsymbol{W}_{\text{PCA},\boldsymbol{\eta}} = \boldsymbol{\Lambda}_{\boldsymbol{\eta}}^{-1/2}\boldsymbol{U}_{\boldsymbol{\eta}}^\top,
\end{align*}
where $\boldsymbol{\Lambda}_{\boldsymbol{\eta}}$ and $\boldsymbol{U}_{\boldsymbol{\eta}}$ are the eigenvalue and eigenvector matrices of the covariance matrix $\text{Var}(\boldsymbol{\eta}) = \boldsymbol{\Sigma}_{\boldsymbol{\eta}}$ such that $\boldsymbol{\Sigma}_{\boldsymbol{\eta}} = \boldsymbol{U}_{\boldsymbol{\eta}}\boldsymbol{\Lambda}_{\boldsymbol{\eta}}\boldsymbol{U}_{\boldsymbol{\eta}}^\top$.

\begin{comment}
\cite{priddle2019efficient}, 
Remarkably \cite{priddle2019efficient} find that the 
need to talk about the warton estimator, and also the type of whitening matrix.
\end{comment}

\section{Transformation KDE in semiBSL}\label{sec:tkdesemiBSL}
Our second contribution significantly improves the robustness of the semi-parametric estimator proposed in \cite{an2020robust} in the context of BSL. As demonstrated in Figure 2, if a given marginal summary statistic distribution has extremely heavy tails, as is common in financial applications for example (see Section \ref{sec:alpha_stable}), the standard KDE does not accurately approximate the true marginal distribution for reasonable sample sizes (number of model simulations in our context). We propose a new semi-parametric estimator that uses transformation kernel density estimation (see \citealp{wand1991transformations}) to model each marginal summary statistic. Like in the classic semiBSL estimator, we model the dependence between the summary statistic dimensions using the Gaussian copula. By doing so, the whitening method proposed in the previous section may be applied in conjunction with the new estimator, to achieve computational gains on top of the improved robustness. In this section, we provide details of our TKDE method for semiBSL. 

Transformation kernel density estimation was introduced by \cite{wand1991transformations}; although, the general ideas have been applied in many different contexts (see, for example,  \citealp{kingma2016improved, parno2018transport}). In brief, the idea is to transform a sample of data so that the standard global bandwidth kernel density estimator (as in (\ref{eq:standard_kde})) is more accurate, and then transform back to the original domain to obtain the estimate of the desired density. 

Recall we are interested in estimating the marginal distributions of the summary statistic vector. That is, for the $j^{\text{th}}$ marginal $\mathcal{S}^{\text{j}}$, we wish to provide an estimate of the true density $g_{S^j}(s)$ with support $\text{supp}(g_{S^j})$ given access to our sample $S(\boldsymbol{x}_1)^j,...,S(\boldsymbol{x}_n)^j$. Hereafter we suppress the $j$ notation for simplicity, and emphasise that we are considering a univariate distribution. Denote a family of bijective and differentiable transformations $\{\mathcal{G}_{\omega} : \omega \in \Omega\}$ indexed by the parameter $\omega$ that map $\text{supp}(g_{S})$ to the real line. The PDF of the transformed random variable $\tilde{\mathcal{S}} = \mathcal{G}_{\omega}(\mathcal{S})$ is given by:
\begin{align*}
	g_{\tilde{S}}(\tilde{s}; \omega) = g_{S}(\mathcal{G}_{\omega}^{-1}(\tilde{s}))\left|\frac{d\mathcal{G}_{\omega}^{-1}(\tilde{s})}{d\tilde{s}}\right|.
\end{align*}
The value of $\omega$ is chosen so that $g_{\tilde{S}}$ is approximately Gaussian. Given this, KDE should provide an accurate approximation of the PDF on the transformed domain according to
\begin{align*}
	\hat{g}_{\tilde{S}}(\tilde{s};h, \omega) = \frac{1}{n}\sum_{i = 1}^n K_h(\tilde{s} - S(\boldsymbol{x}_i)).	
\end{align*}
An estimate of the density on the original domain is then obtained via the inverse transformation:
\begin{align*}
	\hat{g}_{S}(s;h,\omega) = \frac{1}{n}\sum_{i=1}^n K_h(\mathcal{G}_{\omega}(s) - \mathcal{G}_{\omega}(S(\boldsymbol{x}_i)))\left|\frac{d\mathcal{G}_{\omega}(s)}{ds}\right|.
\end{align*}
The above estimator can be thought of as using a location adaptive bandwidth on the original domain. This allows more appropriate smoothing over all features of the density, and often leads to a more accurate density approximation. Variable bandwidth methods, such as those proposed in \cite{loftsgaarden1965nonparametric} and \cite{breiman1977variable} explicitly model the bandwidth as a function of the data. We find (results not shown) for the test distributions considered in this paper, that TKDE performs better.

A non-trivial aspect of applying TKDE to semiBSL is choosing an appropriate family of transformations, and then finding a method of efficiently estimating $\omega$. The most suitable family of transformations is highly dependent on the shape of the data. \cite{wand1991transformations} focus on right-skewed data and use the shifted power transformation; \cite{yang2003improved} use sequential transformations from the Johnson family to estimate the density of a wide range of distributions and \cite{buch2005kernel} use the Champernowne transformation for heavy-tailed data. Our method can be extended to use any of these transformations (among others), but, due to its flexibility, we focus on the hyperbolic power transformation (HPT) introduced by \cite{tsai2017hyperbolic}, which has not previously been used in the TKDE context. The HPT is given by:
\begin{align*}
	\mathcal{G}_{\omega}(s) = \begin{cases} 
      \nu\sinh(\psi_{-} s)\sech^{\lambda_{-}}(\psi_{-} s)/ \psi_{-} & s\leq 0 \\
      \nu\sinh(\psi_{+} s)\sech^{\lambda_{+}}(\psi_{+} s)/ \psi_{+} & s> 0
   \end{cases}
\end{align*}
where $s$ is median centered, $\omega = \{\nu,\psi_{-},\lambda_{-},\psi_{+},\lambda_{+}\}$, $\nu,\psi_{-}, \psi_{+}>0$ and $|\lambda_{-}|,|\lambda_{+}|\leq 1$. $\lambda_{-},\lambda_{+}$ are the power parameters; $\psi_{-},\psi_{+}$ are the scale parameters, and $\nu$ is the normalising constant. By splitting the data either side of the median, the transformation is able to handle bimodal distributions, provided the modes are not well separated. As demonstrated by \cite{tsai2017hyperbolic}, the HPT outperforms other relevant normality transformations for a wide range of distributions.

There are many different optimality criteria possible to determine $\omega$. \cite{wand1991transformations} and \cite{yang2003improved} use asymptotic results based on minimising the mean integrated square error. Here we follow the approach used in \cite{tsai2017hyperbolic} and use maximum likelihood estimation. That is, given we wish to transform the summary statistics such that the global bandwidth KDE will perform well, we target the standard normal distribution $p(\mathcal{G}_{\omega}(s)) = \frac{1}{\sqrt{2\pi}} \exp \{ -\mathcal{G}_{\omega}^2(s)/2\}$ in our transformation. It can be shown that the objective function is given by:
\begin{align*}
	\log p(S(\boldsymbol{x}_i),\dots,S(\boldsymbol{y}_n)|\omega) = \sum_{i = 1}^n \log \phi(\mathcal{G}_{\omega}(S(\boldsymbol{x}_i))) + \log |J(S(\boldsymbol{x}_i))|
\end{align*}
where $\phi$ is the PDF of the standard normal distribution, and the Jacobian term is:
\begin{align*}
	|J(s)| = \left|\frac{\partial \mathcal{G}_{\omega}(s)}{\partial s}\right| = \begin{cases}
	\nu(1-\lambda_{-}\tanh^2(\psi_{-} s))\sech^{\lambda_{-}-1}(\psi_{-} s) & s \leq 0\\
		\nu(1-\lambda_{+}\tanh^2(\psi_{+} s))\sech^{\lambda_{+}-1}(\psi_{+} s) & s > 0.
	\end{cases}
\end{align*}
In practice, there are often several solutions to the score equation, however, only one is the global maximum. \cite{tsai2017hyperbolic} employ the simplex method (see \citealp{nelder1965simplex}) to approximate the MLEs by iteratively optimising each split of the data separately and then perturbing the estimate of the slope parameter according to its MLE:
\begin{align*}
	\hat{\nu} = \left(\frac{1}{n} \sum_{i=1}^n (\mathcal{G}_{\omega}(S(\boldsymbol{x}_i) )^2 \right)^{-1/2}.
\end{align*}
In our implementation, we take a similar approach to \cite{tsai2017hyperbolic} in splitting the data and estimating each of the pairs $\{\psi_{-},\lambda_{-}\}$ and $\{\psi_{+},\lambda_{+}\}$ separately. However, we update the value of $\nu$ using the MLE (using the relevant split of the data) for each evaluation of the likelihood, due to its dependence on the other parameters. We find that this approach works well without having to iteratively maximise the parameters and perturb $\nu$. This is crucial in the context of semiBSL as each iteration of MCMC will involve an estimate of the synthetic likelihood at the proposed parameter value. Of course, our method only serves as an approximation of the true maximum, but as we shall demonstrate, this is sufficient to significantly improve the accuracy of the density estimate over standard KDE. Each marginal summary statistic distribution may be estimated in parallel, meaning the overall additional computational time is small. We use the quantile approach outlined in \cite{tsai2017hyperbolic} to initialise $\omega$ for each optimisation problem. Alternatively, optimal parameters found at previous iterations may be used to inform initial parameter values at subsequent iterations.

Despite the appeal of the HPT, we find that for very heavy tailed data, the transformation is not numerically stable. It is also a non-trivial task to reparameterise the transformation such that it is numerically stable. Therefore, we propose an extension of the HPT that uses a series of $\log$ transformations to first reduce the heaviness of tails, allowing the HPT to subsequently be applied more effectively. The $\log$ transformations do not require any estimation of parameters, and so they add negligible computation time. For positively skewed data with heavy kurtosis, we use the transformation:
\begin{align*}
	\tilde{\mathcal{S}} = \log(1 + \mathcal{S} - \min(S(\boldsymbol{x}_1),\dots,S(\boldsymbol{x}_n)) + \Delta)
\end{align*}
where $\Delta = \min(S(\boldsymbol{x}_1),\dots,S(\boldsymbol{x}_n)) - \boldsymbol{s}_{\boldsymbol{y}}+1$ if $\boldsymbol{s_y} < \min(S(\boldsymbol{x}_1),\dots,S(\boldsymbol{x}_n))$, otherwise $\Delta = 0$. Analogously, we use the following transformation for negatively skewed data with heavy kurtosis:
\begin{align*}
	\tilde{\mathcal{S}} = -\log(1-\mathcal{S} +\max(S(\boldsymbol{x}_1),\dots,S(\boldsymbol{x}_n)) + \Delta)
\end{align*}
where $\Delta = \boldsymbol{s}_{\boldsymbol{y}} - \max(S(\boldsymbol{x}_1),\dots,S(\boldsymbol{x}_n))+1$ if $\boldsymbol{s}_{\boldsymbol{y}}>\max(S(\boldsymbol{x}_1),\dots,S(\boldsymbol{x}_n))$, otherwise $\Delta = 0$. Lastly, for symmetric data with heavy kurtosis we use
\begin{align*}
	\tilde{\mathcal{S}} = \text{sgn}(\mathcal{S})\log(1+\mathcal{S}\text{sgn}(\mathcal{S})).
\end{align*}
When one of the above three $\log$ transformations is applied concurrently with the HPT, we refer to each method as semiBSL TKDE1, semiBSL TKDE2, or semiBSL TKDE3, respectively. SemiBSL TKDE0 refers to semiBSL TKDE without an initial $\log$ transformation, and semiBSL TKDE is the general method of using transformation kernel density estimation for semiBSL. We emphasise that the main purpose of the $\log$ transformations is to transform the data such that the HPT can be accurately computed, not to transform the data to Gaussian. It is the HPTs job to Gaussianise the $\log$ transformed data. Figure 1 shows the estimated density after each step of the TKDE procedure for several test densities with known PDF. Each test density is close to standard Gaussian after applying the HPT (row 3, Figure 1), and the final density estimate is close to the true density (row 4, Figure 1). For our semiBSL TKDE method, we recommend the user perform a number of model simulations at $\boldsymbol{\theta}^0$, visualise the marginal summary statistic distributions and then decide whether or not (and which) $\log$ transformation is necessary.

\begin{comment}

\begin{align*}
Y = \log(1+X-\min(X)+\delta)
\end{align*}
where $\delta = \min (X) - X_0 + 1$ if $X_0 < \min(X)$, otherwise $\delta = 0$. Analogously, we use the following transformation for negatively skewed data with heavy kurtosis:
\begin{align*}
Y = -\log(1-X+\max(X)+\delta)
\end{align*}
where $\delta = X_0 -\max(X) + 1$ if $X_0 > \max(X)$, otherwise $\delta = 0$. Lastly, for symmetric data with heavy kurtosis we use
\begin{align*}
Y = \text{sgn}(X)\log(1+X\text{sgn}(X)).
\end{align*}
Even when the $\log$ transformations are not a necessity, we find that they are able to significantly improve the density estimate.

\end{comment}

To illustrate the efficacy of the proposed density estimation procedure, we perform a simulation study using a wide a range of distributions with known density. This follows directly from the work in \cite{an2020robust}. Specifically, we assume the observed data is drawn from the standard Gaussian distribution $y \sim \phi$, and the summary statistic is given by $S(y) = \sinh\left(\frac{1}{\delta}\left(\sinh^{-1}(y)+\epsilon\right)\right)$ (this is the sinh-archsinh transformation of \citealp{jones2009sinh}). $\epsilon$ and $\delta$ control the skewness and kurtosis respectively. Here we choose the values of $\epsilon$ and $\delta$ to reflect the shapes of densities that arise in practice, for example, in the models of Section \ref{sec:results}. We also consider an observed dataset drawn directly from a bimodal Gaussian distribution, such that $y = 0.5\mathcal{N}(3,1) + 0.5\mathcal{N}(8,1)$ and take $S(y) = y$. For each test density, we estimate the PDF using KDE and TKDE for $n=100$, $n = 500$ and $n = 1000$. For TKDE, we show the results using the most appropriate $\log$ transformation (or lack thereof), see Figure 2. Furthermore, we estimate the total variation distance between the true and estimated PDFs using numerical integration over a grid of parameter values based on 1000 independent replicates of the above procedure. We report the sample mean and standard deviation of the 1000 total variation distances (see Table 1). The total variation between two PDFs $f_1(\boldsymbol{\theta})$ and $f_2(\boldsymbol{\theta})$ is given by $\text{tv}(f_1,f_2) = \frac{1}{2}\int |f_1(\boldsymbol{\theta}) - f_2(\boldsymbol{\theta})|d\boldsymbol{\theta}$.

\begin{comment}
or write as:
\begin{align*}
Y = 
\begin{cases} 
      \log(1+X), & x> 0 \\
      -\log(1-X), & x<0
\end{cases}
\end{align*}
\end{comment}

\begin{figure}[h!]\label{fig:TKDE_process}
	\centering\includegraphics[width = 16cm]{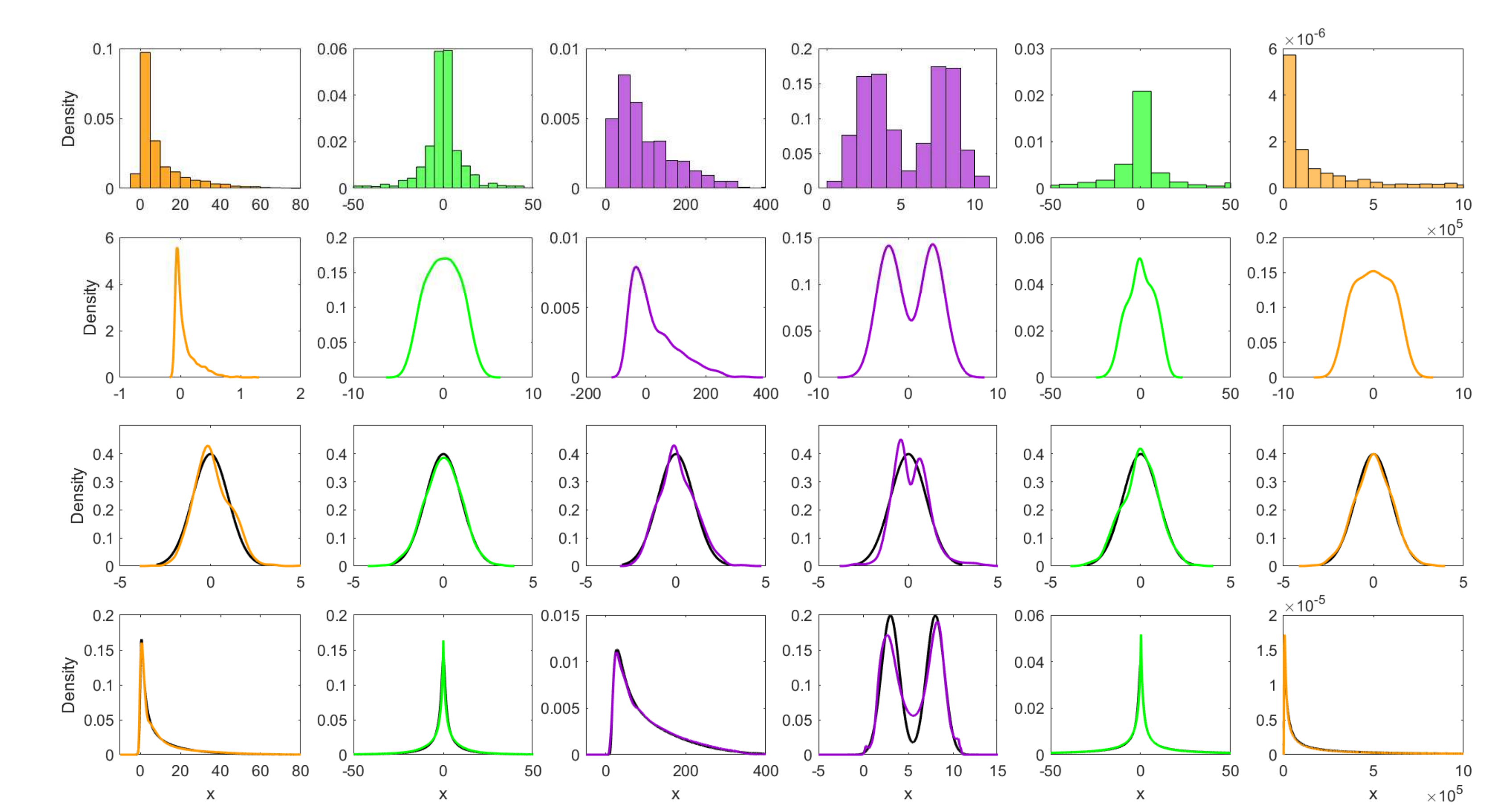}
	\caption{\small Intermediate densities of TKDE procedure for various test densities. Each row corresponds to a step in the density estimation: row 1 is a histogram of the original data; row 2 is a KDE after the log transformation; row 3 is a KDE after the HPT (with the standard normal distribution in black) and row 4 is the final density estimate on the original domain (with the true PDF shown in black). Columns correspond to each test density: skewness and kurtosis ($\epsilon = 1.3, \delta = 0.6$; left), kurtosis only ($\epsilon = 0, \delta = 0.35$), skewness only ($\epsilon = 5, \delta = 1$), bimodal ($0.5\mathcal{N}(3,1)+0.5\mathcal{N}(8,1)$), heavy skewness ($\epsilon = 0, \delta = 0.1$) and skewness with heavy kurtosis ($\epsilon = 5, \delta = 0.4$; right).}
\end{figure}

\begin{figure}[h!]\label{fig:density_estimates}
	\centering\includegraphics[width = 16cm]{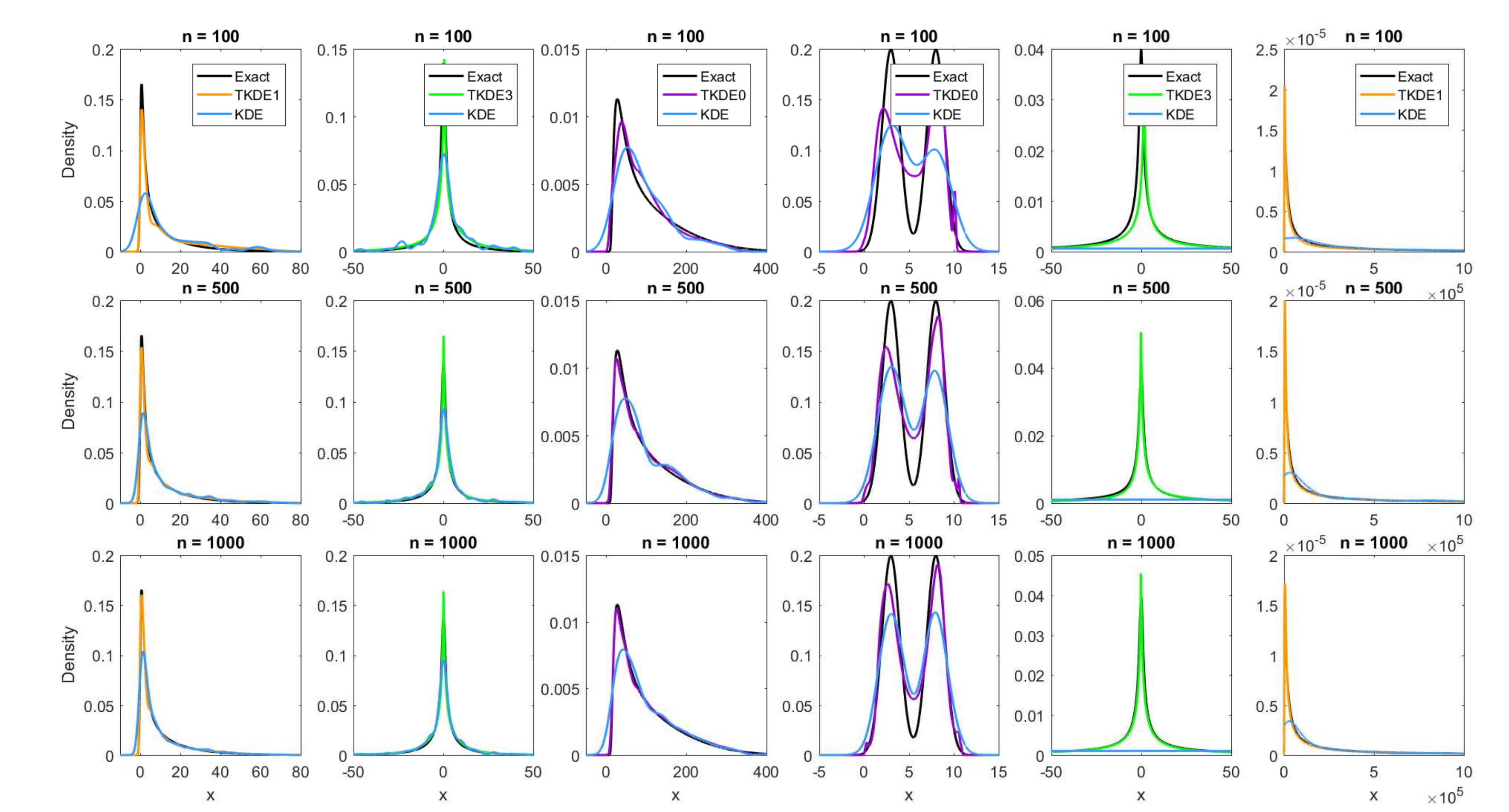}
	\caption{\small Comparison of density estimators (KDE and TKDE) with the true density. Rows correspond to $n = 100$ (top row), $n = 500$ and $n = 1000$ (bottom row) model simulations. Columns correspond to the same test densities listed in Figure \ref{fig:TKDE_process}.}
\end{figure}

%\begin{figure}[tbh]
%	\centering\includegraphics[width = 16cm]{figures/compare_estimators_fig.eps}
%	\caption{\small Same models as before, using $n = 100$ model %simulations. Mean, 0.25 and 0.75 quantiles, using 200 independent replications.}
%\end{figure}

\begin{table}[h!]\label{tab:density_estimates}
\caption{\small Total variation distances between density estimators (KDE and TKDE) and the true density. The mean is reported for each of the four test densities using $n = 100$, $n = 500$ and $n = 1000$ model simulations. The corresponding standard deviations are given in parentheses.
}
\centering
 \begin{tabular}{ p{4cm}|p{1.5cm}p{1.5cm}|p{1.5cm} p{1.5cm}|p{1.5cm}p{1.5cm} }
  \hline
	& \multicolumn{2}{c|}{$n = 100$} &\multicolumn{2}{c|}{$n = 500$} &\multicolumn{2}{c}{$n = 1000$} \\
	\cline{2-7}
	& KDE & TKDE & KDE  & TKDE & KDE & TKDE  \\
	\hline
	\hline
	Skewness and Kurtosis $(\epsilon = 1.3,\delta = 0.6)$ & 0.201 \footnotesize (0.027) & 0.101 \footnotesize (0.033) & 0.138 \footnotesize (0.014) & 0.053 \footnotesize (0.012) & 0.116 \footnotesize (0.010) & 0.041 \footnotesize (0.008) \\
	\hline 
	Kurtosis \hspace{20mm} $(\epsilon = 0, \delta = 0.35)$ & 0.162 \footnotesize (0.022) & 0.095 \footnotesize (0.030) & 0.099 \footnotesize (0.011) & 0.050 \footnotesize (0.011) & 0.079 \footnotesize (0.008) & 0.039 \footnotesize (0.008)\\
	\hline
	Skewness \hspace{20mm}$(\epsilon = 5, \delta = 1)$ & 0.136 \footnotesize (0.023) & 0.072 \footnotesize (0.025) & 0.094 \footnotesize (0.011) & 0.038 \footnotesize (0.011) & 0.080 \footnotesize (0.008) & 0.030 \footnotesize (0.007) \\
	\hline 
	Bimodal \hspace{20mm} $0.5\mathcal{N}(3,1) + 0.5\mathcal{N}(8,1)$ & 0.253 \footnotesize (0.028) & 0.175 \footnotesize (0.032) & 0.189 \footnotesize (0.010) & 0.121 \footnotesize (0.015) & 0.159 \footnotesize (0.007) & 0.100 \footnotesize (0.011) \\
	\hline
	Heavy Kurtosis \hspace{20mm} $(\epsilon = 0,\delta = 0.1)$ & 0.166 \footnotesize (0.013) & 0.058 \footnotesize (0.033) & 0.163 \footnotesize (0.008) & 0.026 \footnotesize (0.011) & 0.165 \footnotesize (0.006) & 0.019 \footnotesize (0.007) \\
	\hline
	Skewness and Heavy Kurtosis $(\epsilon = 5,\delta = 0.4)$ & 0.044 \footnotesize (0.011) & 0.014 \footnotesize (0.009) & 0.023 \footnotesize (0.005) & 0.007 \footnotesize (0.004) & 0.018 \footnotesize (0.004) & 0.006 \footnotesize (0.003) \\
	\hline
\end{tabular}
\label{tab:ar1}
\end{table}

Figure 2 demonstrates that the proposed TKDE scheme is able get much closer to the true PDF than standard KDE, even with a small number of model simulations. The TKDE nicely captures the peaks of each distribution, and provides adequate smoothing over the tails. For $\epsilon = 0, \delta = 0.1$, the KDE appears completely flat due to the extremely heavy tails, whereas the TKDE is very accurate. TKDE also outperforms KDE for the bimodal test density, with a noticeably better performance for $n = 1000$. Interestingly, in some cases, we find that the $\log$ transformation is detrimental to the TKDE, and so the user must carefully decide whether or not the $\log$ transformation is needed. The simulation results in Table 1 support the above findings, with all TKDEs having a lower total variation distance than the corresponding KDE. The benefits of TKDE are most apparent for heavy tailed distributions.

\section{Results}\label{sec:results}
In this section, we apply our methods to four examples. The examples, and what they are designed to demonstrate are listed below:
\begin{comment}
In this section, we apply our methods to four examples. The first example is a simple time-series model with known likelihood, allowing us to compare the result of our methods to the output of a Metropolis-Hastings sampler that uses the true likelihood. We also consider a queueing model, a model from finance and a model from ecology. Each of these models have an intractable likelihood and are representative of a real-life modelling scenario.
\end{comment}

\begin{enumerate}
	\item MA$(2)$ example: demonstrates the potential efficiency gains of wsemiBSL; the robustness of semiBSL TKDE, and the simultaneous use of whitening and TKDE in semiBSL (wsemiBSL TKDE hereafter) for improved effiency and robustness.
	\item Fowler's Toads example: demonstrates the potential efficiency gains of wsemiBSL.
	\item M/G/1 example: demonstrates the improved robustness of semiBSL TKDE.
	\item $\alpha$-stable stochastic volatility model: demonstrates the improved robustness of semiBSL TKDE.
\end{enumerate}
The likelihood for the MA$(2)$ example is known, allowing us to compare the result of our methods to the output of a Metropolis-Hastings sampler that uses the true likelihood. Each of the remaining three models have an intractable likelihood and are representative of a real-life modelling scenario.

% the examples need to be better organised to highlight what each example is trying to demonstrate. i think having a summary at the start of the results section about what each example is aiming to show [would be better] %

In all cases, we use the Metropolis-Hastings algorithm with a Gaussian random walk. The random walk covariance matrix is set to be roughly the (approximate) posterior covariance obtained from pilot runs. Unless stated otherwise, the value of $n$ is tuned such that the standard deviation of the log synthetic likelihood evaluated at $\boldsymbol{\theta}^0$ is in the range $[1,2]$, as \cite{price2018bayesian} find that this maximises the computational efficiency of sBSL. We compare posterior approximations using the total variation distance, as described in Section \ref{sec:tkdesemiBSL}. For wsemiBSL, we use $n_{\text{cov}} = 5000$ to accurately estimate $\boldsymbol{W}$. Each sampler is run for $T = 100000$ MCMC iterations.

\begin{comment}
In this section, we apply the proposed methodology to a series of examples. We demonstrate that semiBSL TKDE can improve the accuracy over standard semiBSL and show that for some models semiBSL TKDE is necessary to obtain a discernible posterior approximation. We also demonstrate that wsemiBSL can significantly reduce the number of model simulations in semiBSL. Furthermore, we show that TKDE can be applied with the whitening transformation for both improved robustness and efficiency over standard semiBSL. We consider four examples: the first example is a toy example with known likelihood, allowing us to reliably assess the performance of the proposed methods, and the final three examples are real-life examples from queuing, finance and ecology. 

\end{comment}

\begin{comment}
we use an MCMC sampler with the random walk set to be roughly the posterior covariance 
we use the same number of model simulations for semiBSL KDE and semiBSL TKDE. We applying shrinkage we allow the number of model simulations to differ.
\end{comment}

\subsection{MA(2)}
The $t^{\text{th}}$ observation $x_t$ in a moving average process of order 2, denoted MA$(2)$, evolves according to:
\begin{align*}
	x_t = w_t + \theta_1w_{t-1} + \theta_2w_{t-2} \hspace{5mm}\text{where}\hspace{5mm} w_i\sim \mathcal{N}(0,\sigma^2) \hspace{5mm}\text{for} \hspace{5mm} i = 1,\dots,T_0
\end{align*}
subject to the constraints $-1<\theta_2<1$, $\theta_1 + \theta_2 > -1$ and $\theta_1 - \theta_2 <1$. It is straightforward to show that the likelihood is Gaussian with zero mean vector and pentadiagonal covariance matrix, with entries given by: $\zeta(0) = 1 + \theta_1^2 + \theta_2^2$, $\zeta(1) = \theta_1 + \theta_1 \theta_2$ and $\zeta(2) = \theta_2$, where $\zeta(k) = \text{Cov}(x_t,x_{t-k})$. The MA$(2)$ model is commonly used as a toy example to demonstrate likelihood-free methods (see, \citealp{chiachio2014approximate,marin2012approximate,nott2019bayesian}).

We simulate 50 observations from the MA$(2)$ process at $\boldsymbol{\theta}_{\text{true}} = (\theta_1,\theta_2)^\top = (0.6,0.2)^\top$ and set this to be our observed data, such that $\boldsymbol{y} = (x_1,\dots,x_{50})^\top$. We assume that $\sigma^2$ is known, and equal to 1. For semiBSL, we are interested in cases where the marginal summary statistic distributions deviate from Gaussian. As in Section \ref{sec:tkdesemiBSL}, we use the sinh-archsinh transformation of \cite{jones2009sinh} to transform and generate a summary statistic with non-Gaussian marginals; thus, $\boldsymbol{s}_{\boldsymbol{y}} = S(\boldsymbol{y})$, where $S(\cdot)$ is the sinh-archsinh transformation applied elementwise. We use a uniform prior over the parameter space.  

We first test our methods with a summary statistic generated with 4 different $\epsilon$ and $\delta$ combinations. We consider $\epsilon = 0, \delta = 1$, which corresponds to no transformation; $\epsilon = -1, \delta = 1$, which creates negative skewness; $\epsilon = 0, \delta = 0.6$, which creates positive kurtosis and $\epsilon = 1, \delta = 2$, which creates negative kurtosis and positive skewness. For each of these summary statistics, we consider the following methods: semiBSL (equivalent to wsemiBSL with $\gamma = 1$); wsemiBSL with $\gamma = 0$; semiBSL TKDE ($\gamma = 1$) and wsemiBSL TKDE with $\gamma = 0$. We compare the results to the `true' posterior, which is obtained using an MCMC sampler with the true likelihood.

\begin{figure}[h!]
	\centering
	\includegraphics[width = 16cm]{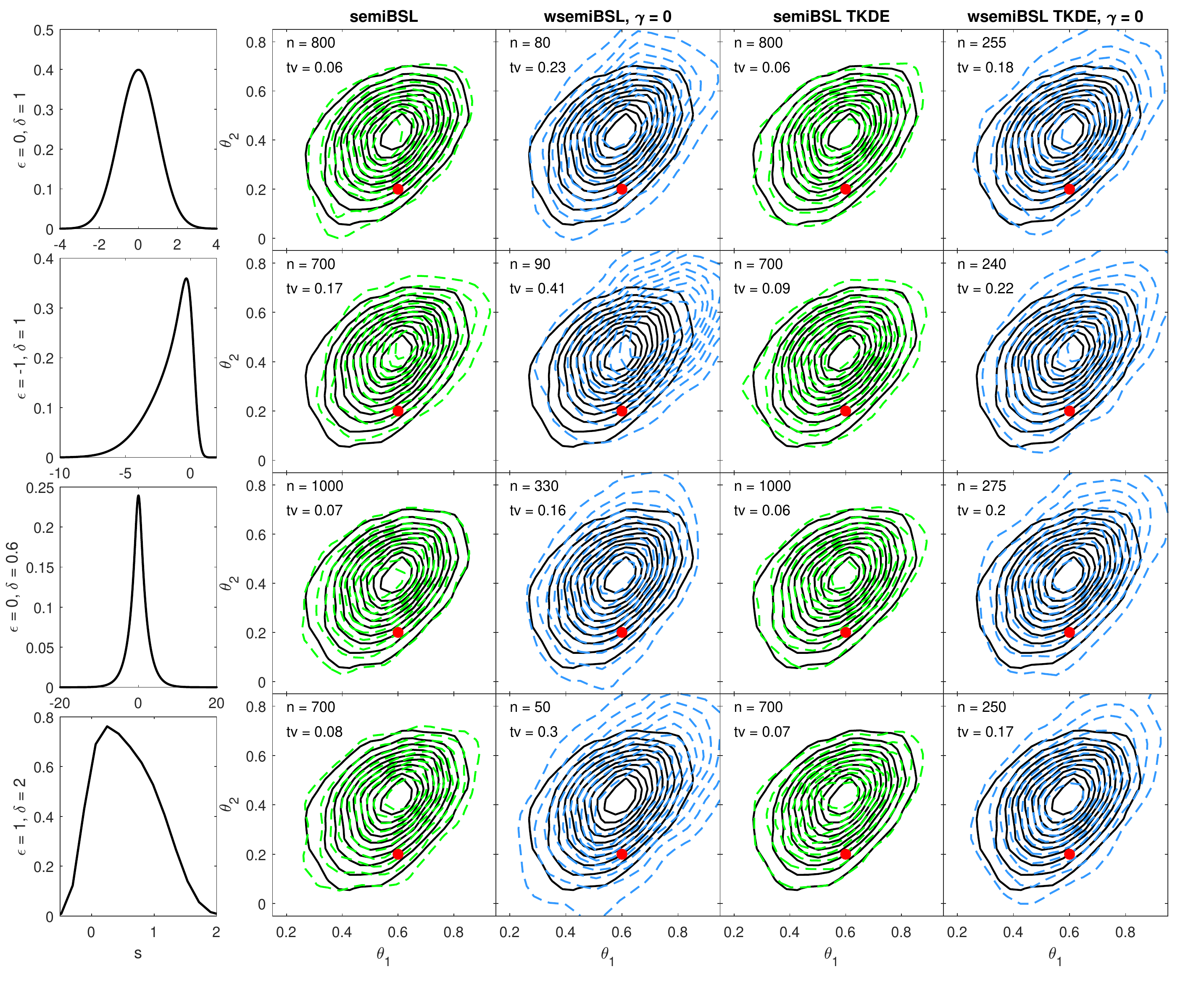}
	\caption{\small Bivariate posterior approximations and true marginal summary statistic distributions for the MA(2) example -- plot 1. Columns denote (left to right) the true marginal summary statistic distribution, semiBSL KDE, wsemiBSL with $\gamma = 0$, semiBSL TKDE and wsemiBSL TKDE with $\gamma = 0$. Each row uses the same marginal summary statistics. Black contours correspond to be the output of an M-H sampler using the known likelihood, green contours are for $\gamma = 1$ and blue contours are for $\gamma = 0$. The parameter used to generate the observed data is shown in red. tv denotes the total variation distance between the approximate and true bivariate posterior distributions.}
		\label{fig:ma2_post}
\end{figure}

Posterior approximations are shown in Figure \ref{fig:ma2_post}. Comparing columns 1 with 3 (no shrinkage, $\gamma = 1$), and columns 2 with 4 (complete shrinkage, $\gamma = 0$), it can be seen that the posterior approximations obtained with TKDE are generally more accurate in terms of the total variation distance to the `true' posterior. 
The only case where the posterior approximation obtained using TKDE is less accurate than the corresponding estimate that uses KDE (albeit slightly, with tv distances of 0.2 and 0.16, respectively), is when $\gamma = 0$ and the marginal summary statistics have negative kurtosis ($\epsilon = 0, \delta = 0.6$; row 3 of Figure \ref{fig:ma2_post}).

The bivariate posterior approximations obtained using wsemiBSL with complete shrinkage are good approximations of the `true' posterior in all cases (Figure \ref{fig:ma2_post}). Interestingly, we find that there is a quite a strong dependence between the regularity of the marginal summary statistic distributions and the capacity of wsemiBSL to significantly reduce the number of model simulations. For no summary statistic transformation, the skewness transformation and the skewness and kurtosis transformation, wsemiBSL is extremely effective -- allowing us to reduce $n$ by about an order of magnitude. However, for the kurtosis transformation, we are only able to reduce $n$ by a factor of three, while accurately estimating the log synthetic likelihood. In addition, we find that wsemiBSL is generally not as effective in reducing the number model simulations when TKDE is used compared to when KDE is used, to estimate the marginal summary statistic distributions. This is the case for all summary statistics except for the kurtosis transformation, where $n$ could be reduced further (from $n = 330$ to $n = 275$) when TKDE is used compared to standard KDE. 

We consider two additional summary statistics with extremely heavy kurtosis. Specifically, we set $\epsilon = 0$ and  $\delta = 0.1$, which creates negative kurtosis, and also $\epsilon = 5$ and $\delta = 0.4$, which creates heavy negative kurtosis and positive skewness. This presents an extremely challenging example for standard semiBSL. We find that $n = 750$ is required for semiBSL TKDE for both datasets and $n = 20000$ is required for semiBSL when $\epsilon = 5$ and $\delta = 0.4$. However, we are unable to find an $n$ that can accurately estimate the $\log$ synthetic likelihood when $\epsilon = 0$ and  $\delta = 0.1$, since the KDE completely fails even for a huge number of model simulations (due to the heaviness of the tails). We also consider $n=750$ for semiBSL, representing the same number of model simulations used for semiBSL TKDE. For these examples, wsemiBSL is ineffective at reducing the required number of model simulations as the marginal summary statistics deviate too far from Gaussian and the pairwise correlation is low. 

\begin{figure}[h!]
	\centering
	\includegraphics[width = 16cm]{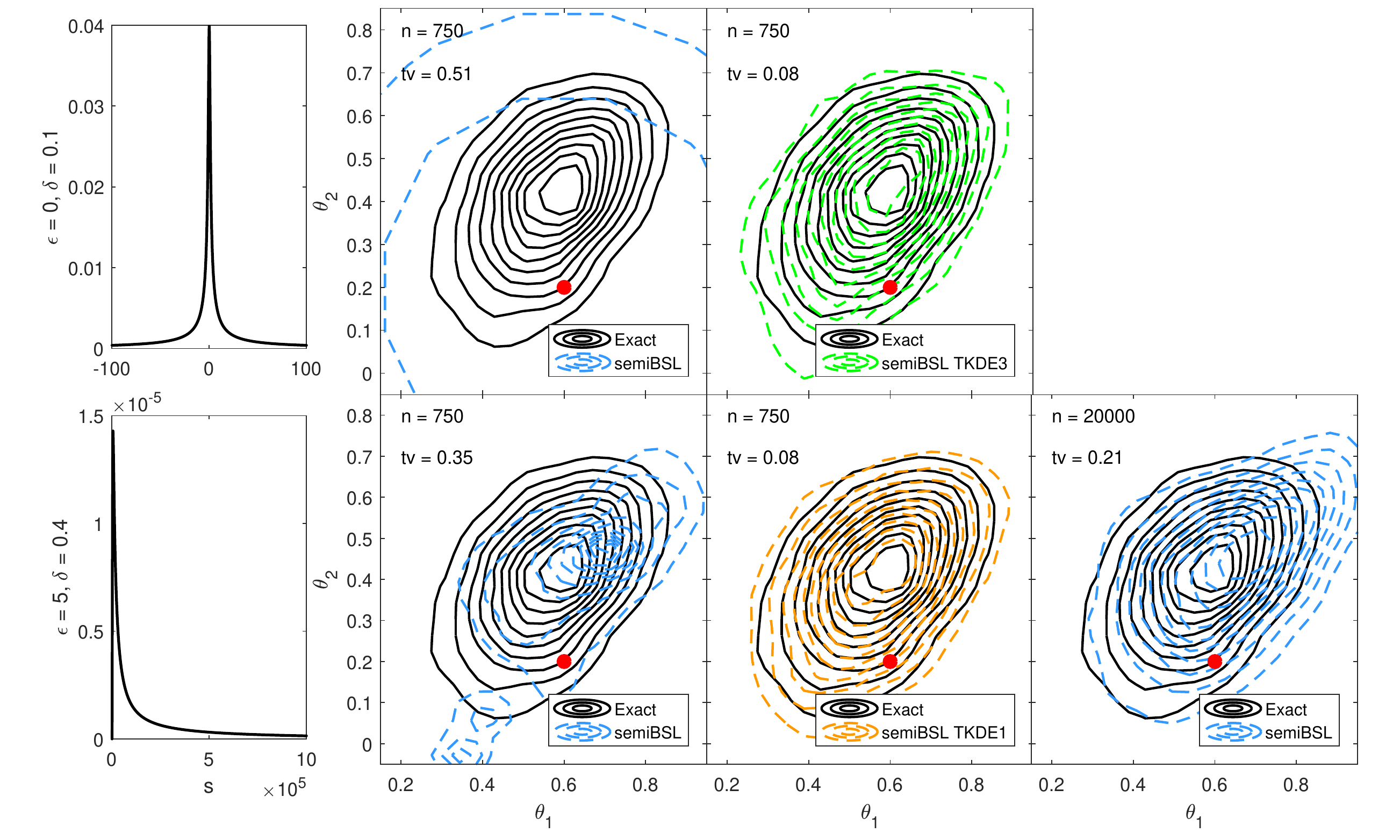}
	\caption{\small Bivariate posterior approximations and true marginal summary statistic distributions for the MA(2) example -- plot 2. Similar to Figure \ref{fig:ma2_post}, the columns (left to right) denote the true marginal summary statistic distributions, semiBSL with $n=750$, semiBSL TKDE with $n=750$ and semiBSL with $n = 20000$.} 
	\label{fig:ma2_tkde_post}
\end{figure}

\begin{figure}[h!]
	\centering
	\includegraphics[width = 16cm]{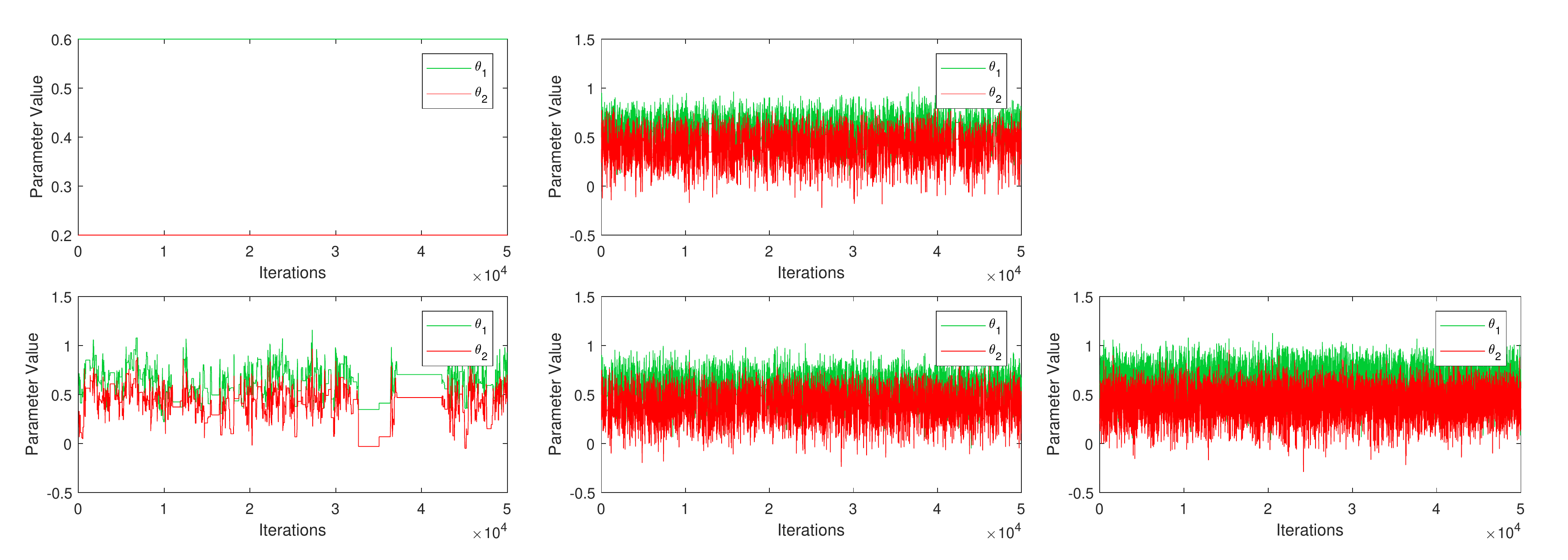}
	\caption{\small Trace plots corresponding to the results in Figure \ref{fig:ma2_tkde_post} (in the respective order). $\theta_1$ is green and $\theta_2$ is red. }
	\label{fig:ma2_trace}
\end{figure}

Bivariate posterior approximations are shown in Figure \ref{fig:ma2_tkde_post}. For $n = 750$, it can be seen that standard semiBSL completely fails, while semiBSL TKDE produces an accurate posterior approximation, for both summary statistics. From Figure \ref{fig:ma2_trace}, it can be seen that the acceptance rates are much higher for semiBSL TKDE than standard semiBSL. For $\epsilon = 0$ and $\delta = 0.1$ for standard semiBSL, the variance of the log synthetic likelihood is so high that no samples are accepted. When $n=20000$ model simulations are used for semiBSL, the parameter space appears to be explored well (Figure \ref{fig:ma2_trace}), but the posterior approximation is far less accurate than the semiBSL TKDE method ($\text{tv} = 0.21$ compared to $\text{tv} = 0.08$), which only used $n=750$ model simulations. 

\subsection{Fowler's Toads}
The next example we consider is the individual-based movement model of Fowler's Toads (\textit{Anaxyrus fowleri}) developed by \cite{marchand2017stochastic}. The model has since been considered as a test example in likelihood-free literature by several authors (see \citealp{an2020robust,frazier2019robust,priddle2019efficient}). \cite{marchand2017stochastic} consider three models, each assuming that toads take refuge during the day and forage throughout the night. The models differ in their returning behaviour; here we expressly consider the random return model. We provide only a brief overview of the model herein, and refer the reader to \cite{marchand2017stochastic} for more details.

To simulate from the model, we draw an overnight displacement from the Levy alpha-stable distribution $S(\alpha, \xi)$, where $0\leq\alpha\leq 2$ and $\xi>0$. At the end of the night, toads return to their previous refuge site with probability $p_0$, or take refuge at their current overnight displacement. In the event of a return on day $i$, the refuge site is chosen randomly from the set of previous refuge sites, thereby giving higher weighting to sites that have been visited multiple times. Here $\boldsymbol{y}$ is the refuge locations of $n_t = 66$ toads over $n_d = 63$ days, generated at $\boldsymbol{\theta}_{\text{true}} = (\alpha,\xi,p_0)^\top = (1.7,35,0.6)^\top$. 

The summary statistic is 48-dimensional, and is constructed as follows. For each toad, we split the observed data in two, corresponding to displacements less than or greater than 10m. We count the number of absolute displacements less than 10m. For the latter dataset, we find the distance moved distribution at time lags 1, 2, 4 and 8 days, and compute both the log of the differences in the $0,0.1,\dots,1$ quantiles and the median. For this example, the marginal summary statistic distributions are roughly Gaussian (see Appendix A, Figure \ref{fig:toad_ss}), meaning sBSL or wBSL would likely perform well. However, semiBSL (and wsemiBSL) will provide additional robustness over their Gaussian counterparts with little additional computation and so we would generally advocate to use these methods even for such models. Of course, TKDE is not necessary for this example.

\begin{figure}[h!]
	\centering
	\includegraphics[width = 16cm]{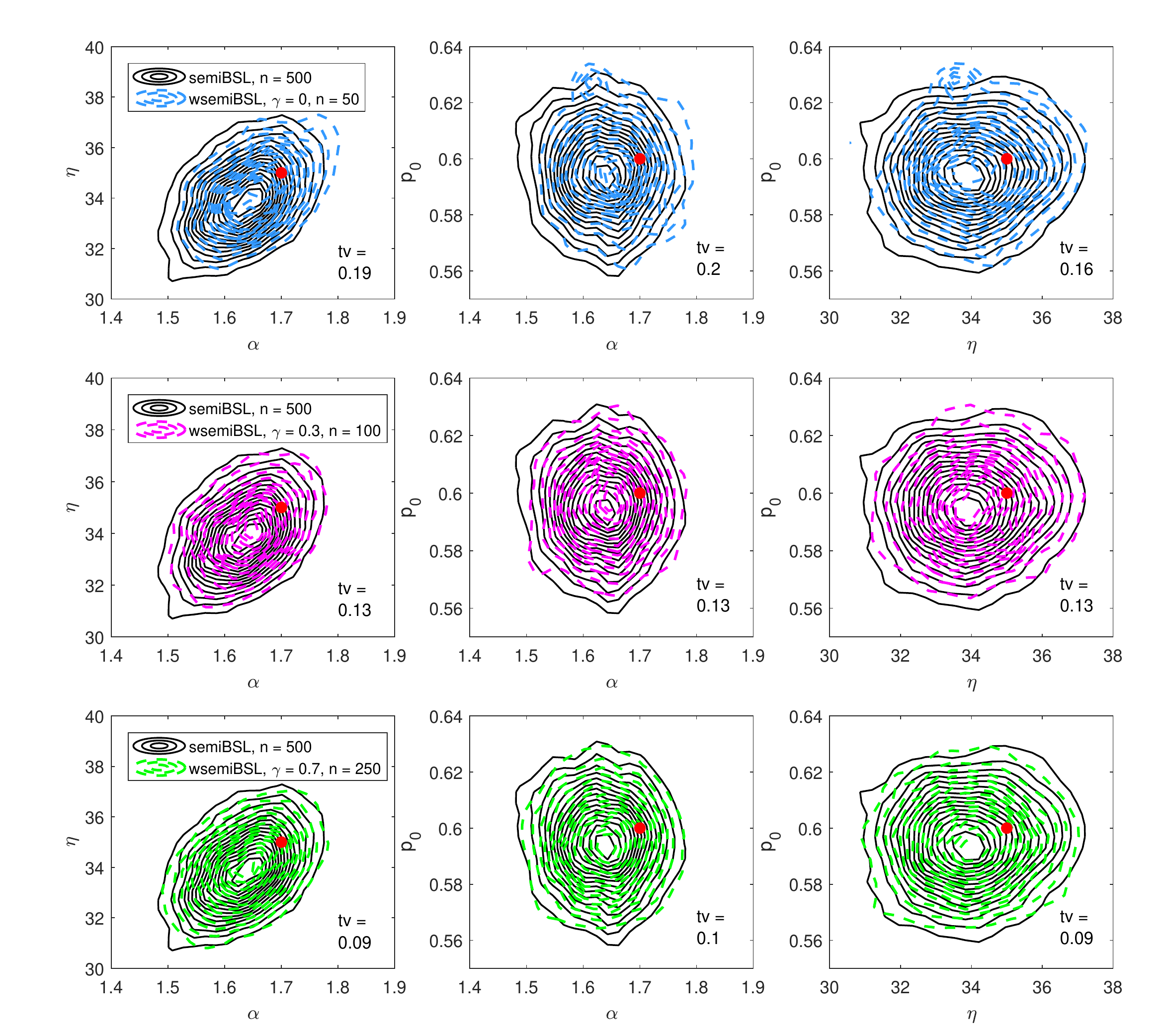}
	\caption{\small Contour plots of the bivariate posterior approximations for the toad model. Rows (top to bottom) correspond to the $n$ and $\gamma$ combination, and columns correspond to each pair of parameters. $\boldsymbol{\theta}_{\text{true}}$ is shown as a red dot. The total variation distance between each bivariate semiBSL posterior approximation and each bivariate wsemiBSL posterior approximation is shown in the bottom right of each panel.}
	\label{fig:toad_post}
\end{figure}

We find that $n = 500$ model simulations is adequate for standard semiBSL. We compare the output of standard semiBSL to wsemiBSL with $n = 250$ ($\gamma = 0.7$), $n = 100$ ($\gamma = 0.3$) and $n  = 50$ ($\gamma = 0$) -- results are shown in Figure \ref{fig:toad_post}. For all cases, the wsemiBSL posterior approximation is close to the standard semiBSL posterior approximation. With complete shrinkage ($\gamma = 0$), we are able to reduce the number of model simulations by an order of magnitude. 

\subsection{M/G/1}
The M/G/1 queueing model is a stochastic single-server queue model whereby `customers' arrive according to a Poisson process and service times have a general distribution. Here we expressly consider the case where service times are $\mathcal{U}(\theta_1,\theta_2)$, as this has been a popular choice in other likelihood-free literature (see e.g.\ \citealp{an2020robust,blum2010non}). The time between arrivals is $\text{Exp}(\theta_3)$ distributed. We assume that only the inter-departure times are known, and take this to be the observed data $\boldsymbol{y}$. We observe 50 inter-departure times (corresponding to 51 customers) and set $\boldsymbol{s_y} = \boldsymbol{y}$, generated at $\boldsymbol{\theta}_{\text{true}} = (\theta_1,\theta_2,\theta_3)^\top = (1,5,0.2)^\top$. The prior is $\mathcal{U}(0,10)\times \mathcal{U}(0,10) \times \mathcal{U}(0,0.5)$ on $(\theta_1,\theta_2-\theta_1,\theta_3)$. 

The marginal summary statistic distributions are right skewed with moderate kurtosis (see Appendix A, Figure \ref{fig:mg1_ss}). Thus, for our TKDE method, it would be reasonable to use semiBSL TKDE0 or semiBSL TKDE1. wsemiBSL does not provide additional benefit for this example since the summary statistics have very low correlation. We run semiBSL TKDE0, semiBSL TKDE1 and standard semiBSL. We compare the results of each sampler to the `true' posterior, obtained using the MCMC scheme for the M/G/1 queue model of \cite{shestopaloff2014bayesian}. We use $n = 500$ to estimate the summary statistic likelihood for semiBSL. 

\begin{figure}[h!]
	\centering
	\includegraphics[width = 16cm]{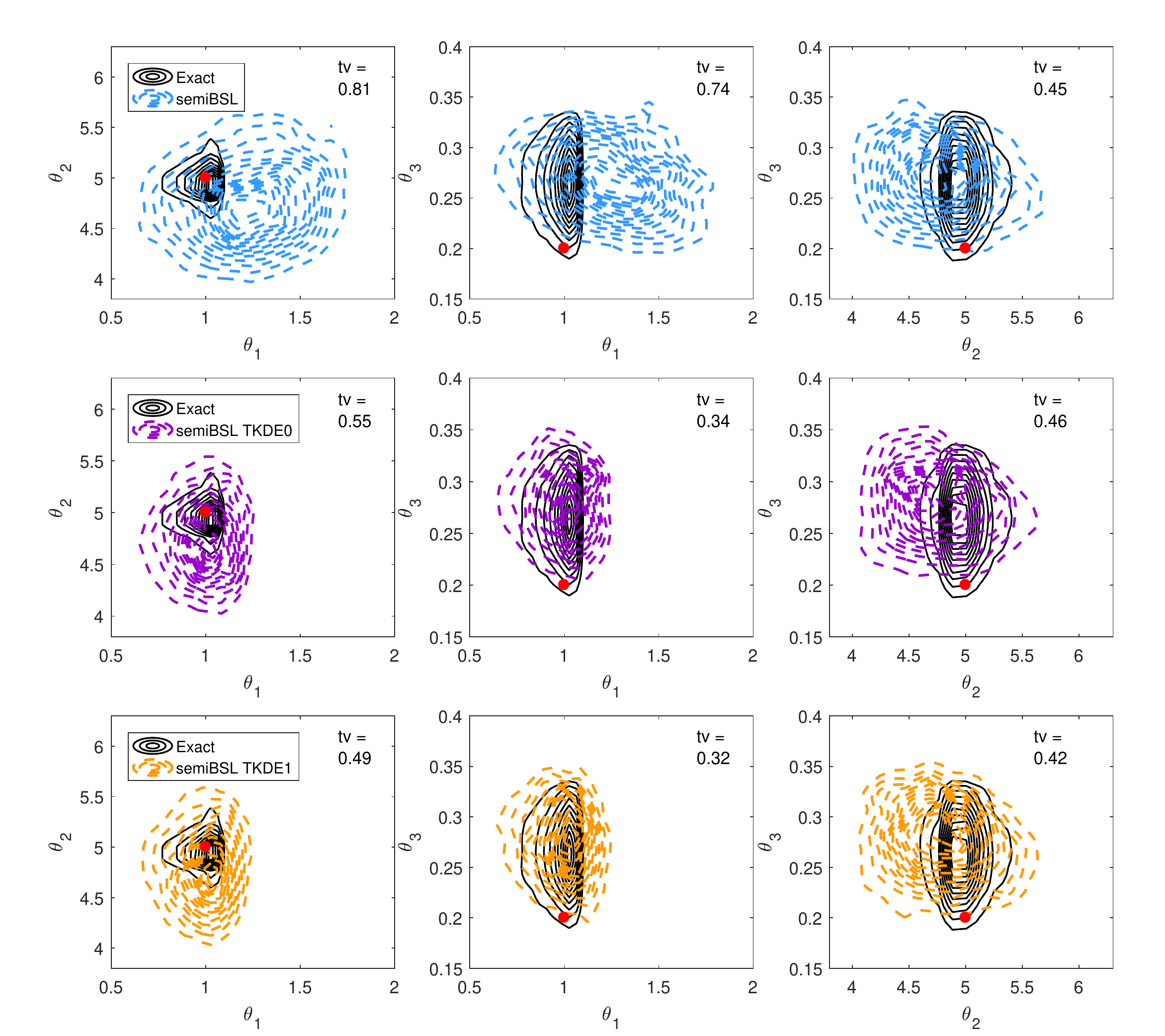}
	\caption{\small Contour plots of the bivariate posterior approximations for the M/G/1 example. Columns (left to right) correspond to each bivariate marginal $(\theta_1,\theta_2)$, $(\theta_1,\theta_3)$ and $(\theta_2,\theta_3)$, respectively. Rows correspond to the method used. `Exact' denotes the posterior approximation obtained using the method of \cite{shestopaloff2014bayesian}. $\boldsymbol{\theta}_{\text{true}}$ is shown as a red dot.}
	\label{fig:mg1_bivariate_posterior_approximations}
\end{figure}

Bivariate posterior approximations are shown in Figure \ref{fig:mg1_bivariate_posterior_approximations}. Both semiBSL TKDE methods produce more accurate posterior approximations than standard semiBSL. semiBSL TKDE estimates the $\theta_1$ marginal distribution more accurately than semiBSL; the $\theta_2$ and $\theta_3$ marginals are similar. The additional $\log$ transformation (in semiBSL TKDE1 compared to semiBSL TKDE0) slightly improves the accuracy of the posterior approximation for this example, as evidenced by the total variation distance. 

\begin{comment}
Acceptance rates of $10.34\%$, $11.8\%$ and $12.42\%$ for KDE, TKDE1 and TKDE, respectively. We use $n = 400$ for each sampler and $M = 100000$ iterations.
\end{comment}

\begin{comment}
\begin{itemize}
	\item the value of $n$ in semiBSL depends heavily on the regularity of the summary statistics. the tKDE can mitigates the dependency of $n$ on the regularity of the summary statistic
\end{itemize}
\end{comment}

\subsection{$\alpha$-Stable Stochastic Volatility Model}\label{sec:alpha_stable}
\begin{comment}
\begin{itemize}
	\item the $\alpha$-stable stochastic volatility models provides a flexible framework for capturing asymmetry and heavy tails, which is useful for modelling financial returns
	\item the $\alpha$-stable lacks a closed form expression for the PDF, which prevents the direct application of standard Bayesian filtering and estimation techniques such as SMC and MCMC
	\item application to propane weekly spot prices
\end{itemize}
\end{comment}
Stochastic volatility models (SVMs) are commonly used in econometric applications, such as the modelling of financial returns (see \citealp{vankov2019filtering}). In SVMs, the observed data are assumed to follow a latent stochastic process in evenly spaced discrete time. The return process is given by:
\begin{align*}
	y_t &= \exp\left(\frac{x_t}{2}\right)v_t\\	
	x_t &\sim \mathcal{N}(\mu + \phi(x_{t-1}-\mu), \sigma_t)
\end{align*}
where $y_t$ is the observed data at time $t$, which directly depends on the log volatility $x_t$ and the shock $v_t$; $\mu$ is the modal instantaneous volatility; $\phi$ is the persistence parameter, and $\sigma_t$ is the variance of $x_t$. The typical model formulation uses a Gaussian shock parameter, $v_t\sim\mathcal{N}(\cdot,\cdot)$ \citep{kim1998stochastic}; however, due to the heavy tailedness of asset returns, more recent studies have found the stable distribution to be more appropriate \citep{casarin2004bayesian}. That is, we assume $v_t \sim \mathcal{SD}(\alpha,\beta,\kappa,\eta)$, where $\alpha$, $\beta$, $\kappa$ and $\eta$ control the tail heaviness (with a lower $\alpha$ having heavier tails), the skewness, the scale and the location, respectively. Despite the additional flexibility inherited by this family of SVMs, the PDF of the stable distribution is unavailable in closed form for most parameter values. This motivates the development of likelihood-free algorithms such as ABC and BSL for heavy tailed distributions (see, for example, \citealp{ebert2019combined,vankov2019filtering}). The extremely heavy tails of $y_t$ may cause sBSL and standard semiBSL to fail. 

\begin{figure}[h!]
	\centering
	\includegraphics[width = 16cm]{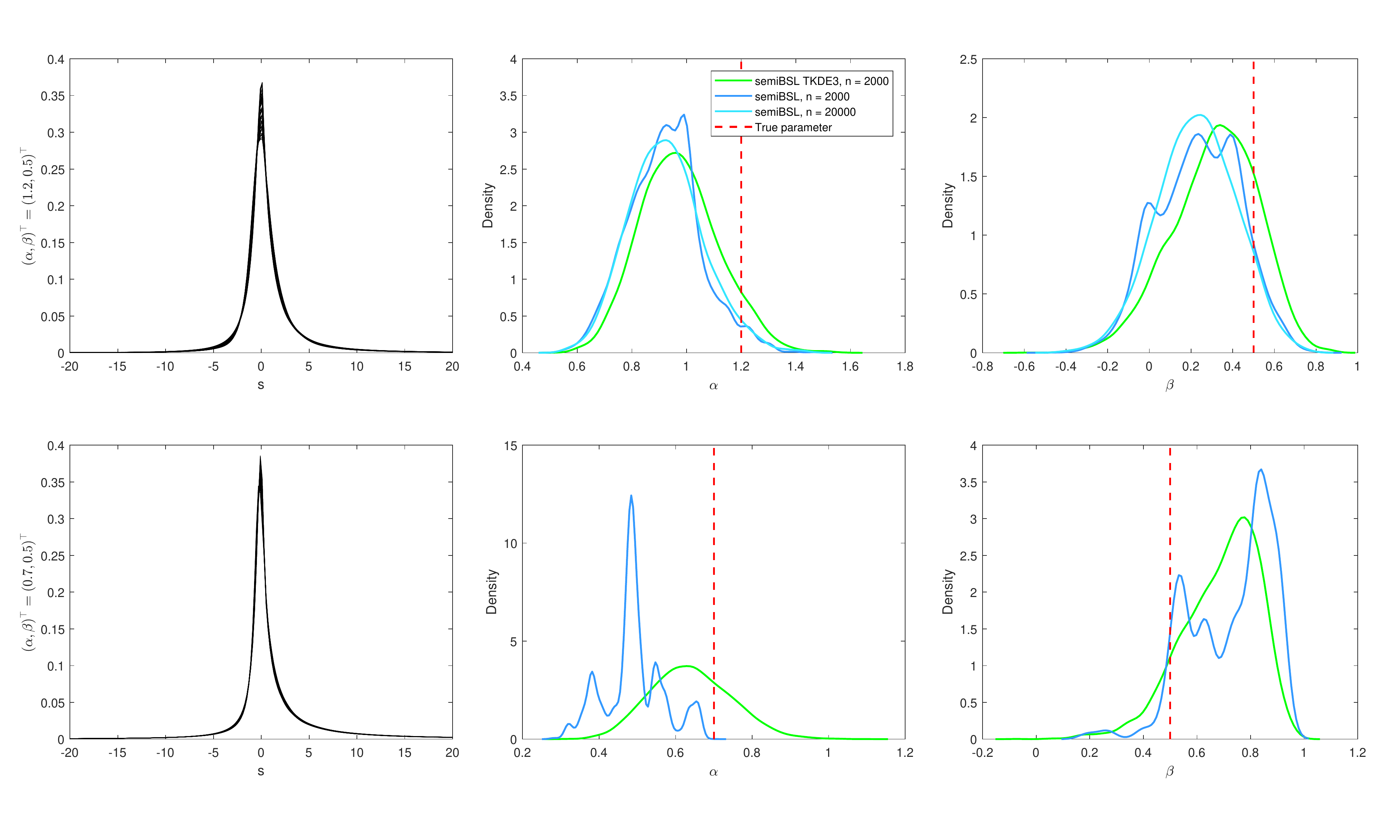}
	\caption{\small Posterior approximations and marginal summary statistic distributions for the $\alpha-$stable SVM. Top row corresponds to dataset 1, and bottom row corresponds to dataset 2. The columns (left to right) correspond to the marginal summary statistic distributions, and the parameters $\alpha$ and $\beta$, respectively.}
	\label{fig:stable_post}
\end{figure}

We test our methods on two datasets. We infer $\boldsymbol{\theta} = (\alpha, \beta)^\top$ and assume the remaining parameters are known and fixed, such that: $\mu = 5$, $\phi = 1$, $\kappa = 1$, $\eta = 0$ and $\sigma = 0.2$ for each dataset. We set $\boldsymbol{\theta}_{\text{true}} = (1.2,0.5)^\top$ and $\boldsymbol{\theta}_{\text{true}} = (0.7,0.5)^\top$ for datasets 1 and 2, respectively and generate 50 observations from the $\alpha-$stable SVM and set this to be $\boldsymbol{y}$ in each case. We take $\boldsymbol{s}_{\boldsymbol{y}} = \boldsymbol{y}$. Given the marginal summary statistic distributions are symmetric and heavily skewed (see Figure \ref{fig:stable_post}), we use semiBSL TKDE3. We do not consider wsemiBSL for this model, since there is only a low degree of correlation between the pairwise statistics. The results are compared directly to standard semiBSL. We find $n = 2000$ is sufficient to control the variance for semiBSL TKDE for each dataset, and $n=20000$ is required for semiBSL for the $\boldsymbol{\theta} = (1.2,0.5)^\top$ dataset. We are unable to find a large enough $n$ to accurately estimate the standard semi-parametric synthetic likelihood for the $\boldsymbol{\theta} = (0.7,0.5)^\top$ dataset. Similar to the MA$(2)$ example, we also consider $n=2000$ for semiBSL for each dataset -- the same value of $n$ we use for semiBSL TKDE. 
%
%We tune $n$ with respect to semiBSL TKDE, and use the same value for semiBSL. We find $n = 2000$ is sufficient to control the variance of the $\log$ synthetic likelihood. 

\begin{figure}[h!]
	\centering
	\includegraphics[width = 16cm]{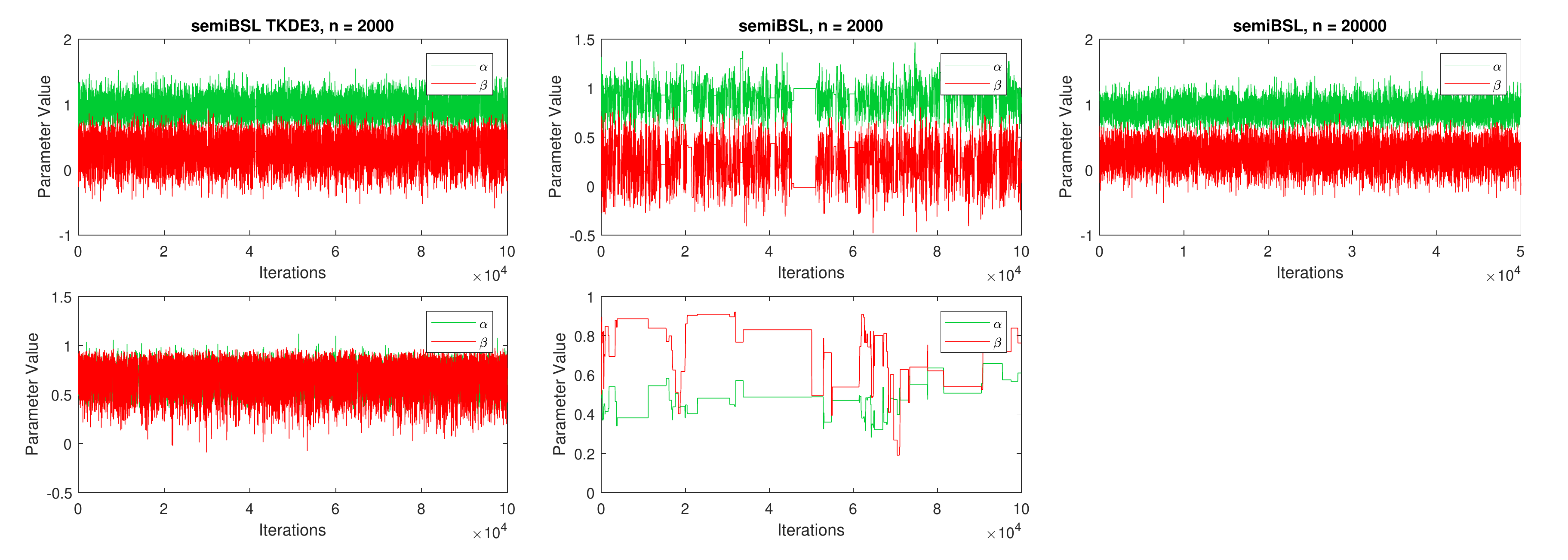}
	\caption{\small Trace plots corresponding to Figure \ref{fig:stable_post} results. Rows correspond to each dataset -- top row when $\boldsymbol{\theta} =(1.2,0.5)^\top$ and bottom row when $\boldsymbol{\theta} = (0.7,0.5)^\top$. Columns correspond to the method and number of model simulations combination.}
	\label{fig:stable_trace}
\end{figure}

Marginal posterior approximations are shown in Figure \ref{fig:stable_post}. The corresponding trace plots are shown in Figure \ref{fig:stable_trace}. We observe similar results to the MA$(2)$ example. For $n=2000$, for standard semiBSL, the acceptance rate is low (extremely low for dataset 2), while we observe high acceptance rates and good mixing for semiBSL TKDE for both datasets. The posterior approximations for dataset 1 obtained using semiBSL are reasonable, but are poor for dataset 2. On the other hand, the posterior approximations for semiBSL TKDE for each dataset are smooth and have reasonable support for the true parameter value. When $n$ is increased to 20000 for standard semiBSL, the posterior approximation is smoother; however there is less support for the true parameter value than the posterior approximation generated using semiBSL TKDE. 

\section{Discussion}
In this article, we proposed two extensions to semiBSL. First, we extended the wBSL method of \cite{priddle2019efficient} to the semiBSL context. We demonstrated in a number of empirical examples that our new method, wsemiBSL, is able to produce accurate posterior approximations with up to an order of magnitude less model simulations than standard semiBSL, even when the summary statistic deviates from normality. We also proposed a new method to estimate the marginal summary statistic distributions in semiBSL using TKDE. We show several examples where standard semiBSL will fail due to heavy kurtosis in the marginal summary statistic distributions, whereas our semiBSL TKDE method produces accurate posterior approximations in each case. Furthermore, we showed that wsemiBSL can be used in conjunction with TKDE for both improved computational efficiency and robustness to irregular summary statistic distributions. 

There are a few limitations to the proposed methods. For wsemiBSL, we find that there is a rather strong dependence between the regularity of the marginal summary statistic distributions and the potential for large reductions in $n$. That is, the efficiency gain appears to diminish as the marginal summary statistic distributions become increasingly non-Gaussian.  

In future work, it may be of interest to consider ways to further increase the robustness of semiBSL to non-linear dependence structures. One way of overcoming such a problem may be via more advanced multivariate transformations such as normalising flows (see \citealp{rezende2015variational,papamakarios2017masked}). In addition, sBSL is known to be adversely affected in the setting of model misspecification, or more specifically, summary statistic incompatibility (see \citealp{frazier2019robust}). Future work may investigate the equivalent problem in the context of semiBSL.

\begin{comment}
 This leads to the following formulation of the SVM: 
\begin{align*}
	y_t &= \exp\left(\frac{x_t}{2}\right)v_t\\	
	x_t &\sim \mathcal{N}(\mu + \phi(x_{t-1}-\mu), \sigma_t)
\end{align*}
where $y_t$ is the observed data at time $t$, which directly depends on the log volatility $x_t$ and the shock $v_t$; $\mu$ is the modal instantaneous volatility; $\phi$ is the persistence parameter, and $\sigma_t$ is the variance of $x_t$. Each $v_t$ is drawn independently across $t$ from the stable distribution,Despite the additional flexibility inherited by this family of SVMs, the PDF of the stable distribution is unavailable in closed form. A number of authors have considered ABC methods to overcome , and so of the stable distribution is $v_t$ is drawn independently across $t$ from the stable distribution. 

\end{comment}

\begin{comment}
\subsubsection{AR(1)}
(need to update these results)
\begin{figure}[H]
	\centering
	\includegraphics[width = 16cm]{ar1_res.pdf}
	\caption{\small Posterior approximations for the AR(1) example. Whitening and tkde results.}
\end{figure}
\end{comment}

\bibliographystyle{apalike} 
\bibliography{BIB} 

\newpage
\appendix
\section{Summary Statistic Distributions}
\begin{figure}[h!]
	\centering
	\includegraphics[width = 16cm]{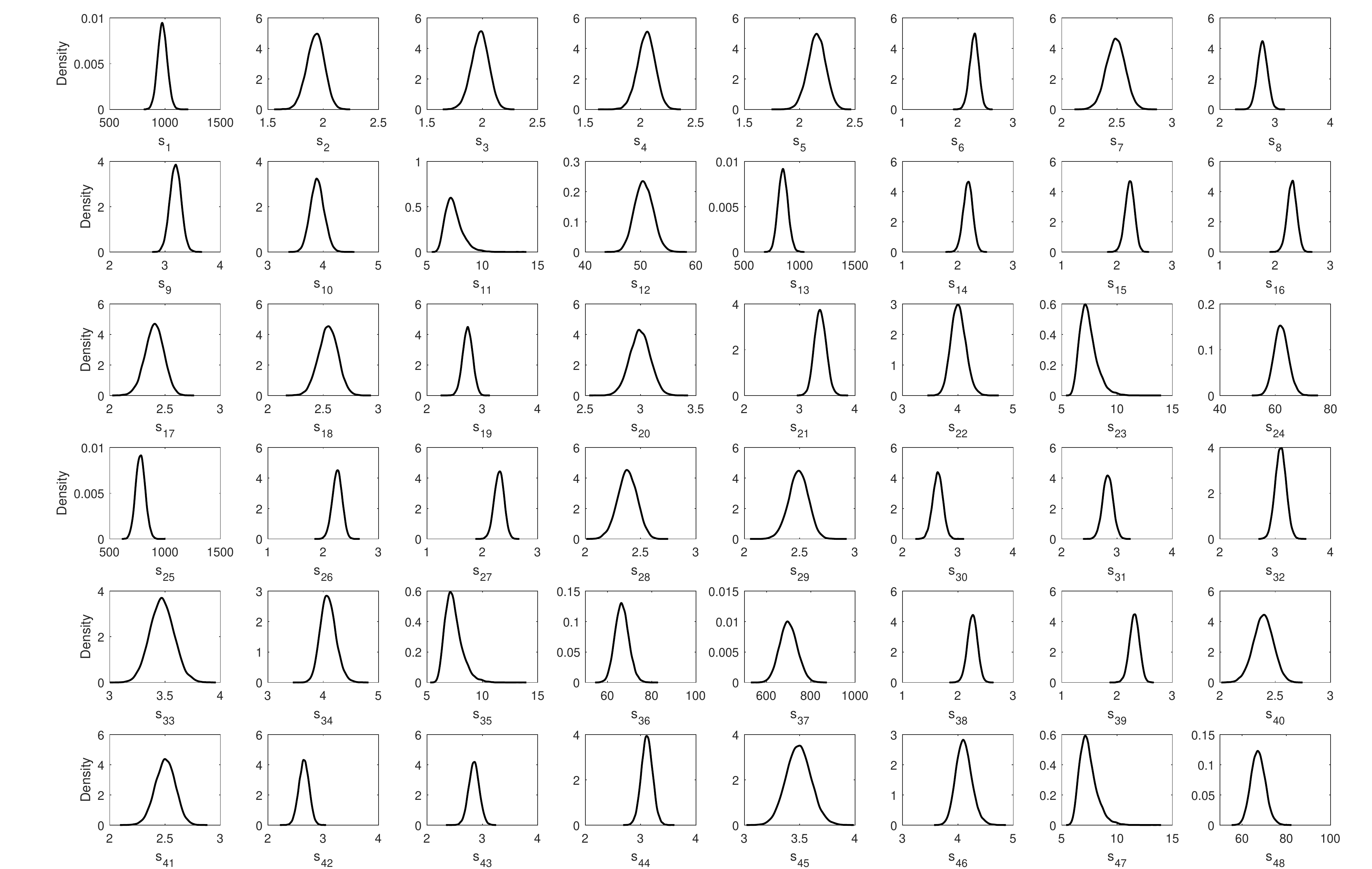}
	\caption{\small Marginal summary statistic distributions for the Fowler's toads example.}
	\label{fig:toad_ss}
\end{figure}

\begin{figure}[h!]
	\centering
	\includegraphics[width = 16cm]{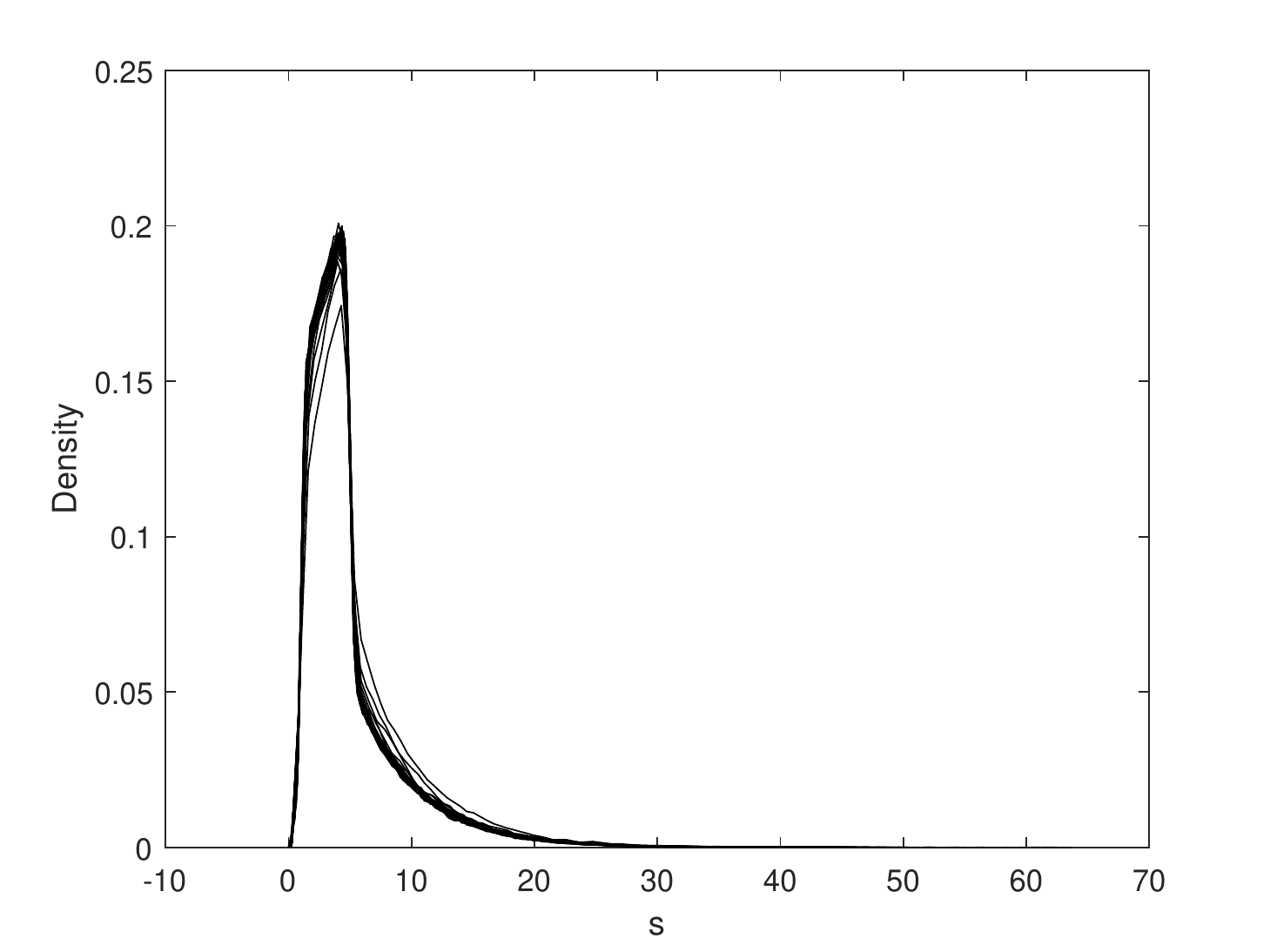}
	\caption{\small Marginal summary statistic distributions for the M/G/1 example.}
	\label{fig:mg1_ss}
\end{figure}

%%%%%%%%%%%%%%%%%%%%%%%%%%%%%%%%%%%%%%%%%%%
%%%%%%%%%%%%%%%%%%%%%%%%%%%%%%%%%%%%%%%%%%%

%%%%%%%%%%%%%%%%%%%%%%%%%%%%%%%%%%%%%%%%%%%
%%%%%%%%%%%%%%%%%%%%%%%%%%%%%%%%%%%%%%%%%%%

\end{document}